\documentstyle[epsfig]{mn} 
\input epsf 
 
\title[Search for star clustering]
{Search for star clustering: methodology and application to the 
Two Micron Galactic Survey} 
  
\author[M. L\'opez-Corredoira et al.]
       {M. L\'opez-Corredoira$^*$, F. Garz\'on,
        P. L. Hammersley, T. J. Mahoney\\
	Instituto de Astrof\'\i sica de Canarias, E-38200 La 
Laguna, Tenerife, Spain \\
        $^*$ E-mail address: martinlc@iac.es.} 
\date{Accepted xxxx. 
      Received xxxx; 
      in original form xxxx} 
  
\begin{document} 
 
\maketitle 

\begin{abstract}
A new approach to the study of the large-scale stellar cluster
distribution in the Galaxy based on two-point correlation
techniques is presented. The basic
formalism for this method is outlined and its
applications are then investigated  by the use of a simple   model of cluster
distribution in
the Galaxy. This provides an  estimate of  the potentials of the two-point
correlation function for indicating clustering in the measured
star positions, which can be related to the presence of star clusters in
the observed field. This technique is then applied to several 
areas of the Two Micron Galactic Survey catalogue, from which information is obtained on the
distribution of clusters  according to position in the Galaxy, as well as 
 about age,
density of stars, etc.
\end{abstract}
\begin{keywords} 
stars: statistics -- infrared: stars -- 
open clusters and associations: general
-- Galaxy: structure
\end{keywords} 
 
\section{Introduction} 

Open  star clusters provide  
valuable information on the 
evolution of the Galaxy.  
In this paper two-point correlation techniques are used to analyse
the distribution of open clusters in order to  gain an insight into 
the structure and evolution of the Galaxy.

Open-cluster distributions have been widely studied at optical
wavelengths as a means 
of studying Galactic structure and evolution  
(see, for example, Lyng\aa \ 1987b; Janes \& Phelps 1994). 
The Lyng\aa\ catalogue of open clusters  
(Lyng\aa \ 1987a) lists about 1200 clusters,  which represent nearly   
all the open clusters accessible in the  visible. 
Knowledge of the positions and ages of these clusters (a method of  
age determination for clusters is given by Carraro \& Chiosi 
1994) enables the scale length and scale height of the disc to be derived  
for both young and old clusters (for a review of old open clusters, see Friel 
1995) and theories to be developed on their formation and destruction 
throughout the history of the Galaxy.  
 
The limitations on these studies are imposed by the 
maximum distance at which open clusters can be detected.  
Most of the cataloged open clusters are in the solar  
neighbourhood,  and very few  have distances greater than  
3 kpc  
(Payne-Gaposchkin 1979). Hence, information is  
obtained only for a small region of the Milky Way. The problem is caused 
by interstellar extinction.  An excellent tool for studying star clusters and star formation 
regions is to observe in the infrared (Wynn-Williams 1977), where the effect of the extinction
is far less. To date, however, the infrared  has been little used in 
this field owing
the absence of suitable databases. 
 
The $K$ band is probably the best region of the spectrum 
for tracing the stellar distribution of the Galaxy. The $K$ radiation
is a mass tracer in spiral galaxies because it follows the old
stellar population (Rhoads 1995). Furthermore, 
the $K$ light is dominated by high-mass stars  in star formation regions, 
i.e. in open clusters, so it is especially appropriate in the search for 
young clusters, which are rich in massive stars.  

As explained in more detail by Garz\'on et al. (1993), the Two micron 
Galactic survey (TMGS) is a $K$-band survey of various regions along the
Galactic equator between  $-5^o<l<35^o$, $|b|\le 15^\circ$ and $35^\circ<l<180^\circ$,
 $|b|\le 5^\circ$. The TMGS 
catalogue has a positional accuracy of about 4 arcsec in right
ascension and 7 arcsec in declination. These errors have been
estimated after cross-correlating the original TMGS  source positions
with  {\it Guide Star Catalogue} (GSC) counterparts. The larger error 
in declination comes from the orientation of the array with respect to the survey direction.
Due to the dead spaces between
detectors, the $K$ limiting magnitude for completeness has to be set
conservatively within a range from 9 to 9.5 mag, although the
detection limit magnitude of the survey is well in excess of 10 mag.

In this paper a new method is presented for automatically determining
the level of clustering in catalogues,  the TMGS being used as an example.
A set of  criteria are defined for an automatic search for correlations 
among stars by means of  the  
two-point correlation function and the two-point angular correlation 
function.
A simple model which assumes  a regular distribution of clusters 
with a constant star density  is developed in Section 3. Predictions  
from this model are then compared with the TMGS and the discrepancies 
analysed. The causes of clustering are then discussed. 
The use of the tools described in Section 2 and their application to the TMGS
catalogue are dealt with in Section 5, and   
the results for several regions of our Galaxy are given in Section 6.
Finally,  a summary of the results is given, and  
suggestions are made for future developments of the methodology  
described here. 
  
\section {The two-point correlation function and the two-point angular 
correlation function} 
 
Occasionally, when the scale length of a system is changed, 
certain aspects of the system remain invariable, as 
in the case for the distribution of matter in space. 
For example, there are mathematical methods for
handling the spatial distribution of atoms in solids, 
gases and (particularly) liquids. 
 Cosmologists face the same kind of mathematical problem when
working with the distribution of galaxies and clusters  
of galaxies in the context of the large-scale structure of the Universe.
They treat the Universe as a fluid whose `particles' 
are galaxies. Our aim in this paper is to develop the
use of similar mathematical methods
on an intermediate scale, i.e. in examining 
the distribution of the stars that make up our Galaxy.  

Correlation functions describe how points  
are distributed (e.g. Peebles 1980; 
Borgani 1995).  
Suppose  a local density of objects  
$\delta N/{\delta V}=\rho(\vec{r})$ and an average density $\langle\rho\rangle$ 
(hereafter, the symbol $\langle\cdots\rangle$ indicates a local volume average).  Note that 
$\rho $ corresponds to the exact distribution of objects, i.e. it is a Dirac 
delta function with  zero value where there are no objects,  
and $\langle\rho\rangle$ is the 
local average density, i.e. the number of objects per unit volume, and
provides no information concerning 
their distribution.  

The two-point correlation function (TPCF) is defined as 
 
\begin{equation}\xi (|\vec{r}-\vec{r}'|)=\frac{\langle\rho(\vec{r}) \rho(\vec{r}')\rangle} 
{\langle\rho(\vec{r})\rangle^2} -1 
.\label{xi}\end{equation}

The function $\xi (r)$ expresses the excess over the random probability of 
finding objects at 
separation $r$. ($\xi (r)=0$ means that the probability is totally random;  
$\xi (r)>0$  that the probability is greater than random, i.e. that there is 
clustering; and   $\xi (r)<0$ that the probability is less than random, i.e. 
that there is relative avoidance). 
 
In the same way, the two-point correlation 
function can be defined for two  dimensions on the surface on to which the 
distribution is projected (the celestial sphere in the case considered here).
 This is called 
the two-point angular correlation function (TPACF) and is defined as 
 
\begin{equation}\omega (|\vec{\theta}-\vec{\theta}'|)=\frac{\langle\sigma 
(\vec{\theta})  
\sigma(\vec{\theta}')\rangle} 
{\langle\sigma(\vec{\theta})\rangle^2} -1 
 \label{omega},\end{equation} 
where $\sigma $ is the surface density per unit solid angle. 

Another mathematical technique for deciding whether a 
distribution is non-Poissonian is area tessellation, as was
used by Bal\'azs (1995) to 
test the grouping tendency of H$_\alpha$-emission stars 
 in the Orion molecular clouds without giving a quantitative  
measure of the departure from the Poissonian distribution. See also
P\'asztor et al. (1993), P\'asztor \& T\'oth (1995) and references therein
for other astronomical applications of spatial statistics.

\subsection{Relationship between the TPCF and the TPACF for stars} 

When applying the above definitions to stars in the Galaxy, the luminosity
function and space density have to be taken into account.

By generalizing the result of the Limber (1953) equation for constant
density, the relationship between the TPCF (which is non-zero for
distances less than $\Delta r$) and the TPACF (for small $\theta$) for any density distribution is
 
\[ 
\omega _{\rm t}(\theta)\approx \frac{1}{\langle\sigma _{\rm t}\rangle^2} \int _0^{\infty }dr\ r^4 \langle\rho 
\rangle^2(r) 
\int_{r-\Delta r}^{r+\Delta r}dr' 
\]\[ \times
\int _{M_{\rm min}(r)}^{M_{\rm max}(r)} dM\ \phi(M)\int_{M_{\rm min}(r)}^{M_{\rm 
max}(r)}dM'\ 
\phi(M')
\]\begin{equation} \times \xi \left(\sqrt{ 
(r\theta )^2+(r-r')^2};r,M,M'\right) 
,\label{omegat2}\end{equation}
where $r$ is the distance along the line of sight, 
$M$ the absolute magnitude, 
$\phi (M)$ the lu\-mi\-no\-si\-ty function and

\begin{equation} 
\langle\sigma _{\rm t}\rangle=\int _0^{\infty }dr\ r^2\langle\rho \rangle(r)\int _{M_{\rm min}(r)} 
^{M_{\rm max}(r)} dM\ \phi(M) 
.\label{sigmat2}\end{equation} 
The minimum and maximum values of $M$ for a distance $r$ depend on
the minimum and maximum values of the apparent magnitude and the extinction
along the line of sight.
 
In this case, it is assumed that the absorption is not patchy, i.e. that it
is independent of $\theta $ for small angles. This is not exactly true
but it will be  show in Section \ref{.extincorr} that the effects are
negligible.

The subscript `t' stands for
`total', a projection of all distances and magnitudes,
and $\langle\sigma _{\rm t}\rangle$ is the total two-dimensional density for all distances and 
magnitudes.
In the literature, $\langle\sigma _{\rm t}\rangle$ is also called
$A(m_{\rm min},m_{\rm max},l,b)$ and represents the
star counts in the magnitude range (Bahcall 1986). 

This expression enables the TPACF to be found once   
the three-dimensional distribution of the stars is known and forms the basis of this 
article, in which  
we create a model distribution of the stars and compare the 
results obtained with those observationally 
in order to investigate the distribution
of clustering in the structure of our Galaxy.

 In general, the TPACF cannot be inverted to give the TPCF 
due 
to the multiplicity of possible solutions and to the lack of precise knowledge 
of certain parameters. However,
there are certain cases in which the equation can be inverted and 
TPCF obtained from TPACF (Fall \& Tremaine 1977). 
A trivial example where inversion is possible is that of a 
Poissonian three-dimensional distribution, which
implies a Poissonian projected distribution and vice versa, i.e. $\xi =0$, 
$\omega =0$ on all scales. Another example is when  $\langle\rho\rangle(r)$ is a constant
independent of $r$.   
  
\subsection{ Definition of new variables} 

In order to simplify the comparison of the level of clustering for
different regions of the sky, two new variables will be introduced.
 
$\theta _{\rm max}$ is defined as the first zero of $\omega _{\rm t}(\theta)$. 
In this article (see for example Fig. \ref{Fig:a052}), $\omega _{\rm t}$ is positive 
up to a separation $\theta _{\rm max}$. For values greater 
than $\theta _{\rm max}$ this value is small and oscillates about zero,
as there is no correlation among stars separated by
large angular distances. 

Another definition, corresponding to the integration of $\omega _{\rm t}$ up
to the limit $\theta =\theta _{\rm max}$ ($\theta >\theta _{\rm max}$   
would give a null contribution to the integral), is
 
\begin{equation}  
C_2\equiv \frac{\int _0 
^{\theta _{\rm max}}d\theta \ \theta \omega _{\rm t}(\theta ) }{\theta _{\rm max}^2} 
,\label{C2}\end{equation} 
which means the excess
(when $C _2$ is positive) or deficit (when $C _2$ is negative) of the relative
number of objects
with respect to a Poissonian distribution 
in a circle centred on an arbitrary star on the celestial sphere,
within the observed solid angle and with angular radius 
$\theta _{\rm max}$. The relative  correlation within the
angular scale $\theta _{\rm max}$ is therefore measured.
We call $C_2$ the `accumulation parameter' 
(N.B. there are also other definitions in the literature of the
TPACF integral, e.g. Wiedemann \& Atmanspacher 1990).

The variable $C_2$ has a clear meaning associated with projected clustering 
and is also a useful number to measure. Since
it sums several values of $\omega $ for different angles, it condenses
the information of interest into a single number that can be compared  
for different samples of stars and give the degree of clustering.   
This parameter is a mathematical expression of the degree of clustering 
seen in fields of stars. 
The idea that we wish to stress here is that all mathematical  
developments  described in this paper are designed to put the intuitive idea 
of clustering to a reliable test. 
These calculations are necessary for a quantitative, 
as opposed to a merely qualitative, description of clustering. 
 
Applying the expression (\ref{omegat2}) of $\omega _{\rm t}$ in $C_2$, we get 
  
\[ 
C_2=\frac{1}{\langle\sigma _{\rm t}\rangle^2\theta _{\rm max}^2} 
\int _0^{\infty }dr\ r^2\langle\rho (r)\rangle^2 
\int _{M_{\rm min}(r)}^{M_{\rm max}(r)} dM\ \phi(M) 
\]\begin{equation} \times
\int _{M_{\rm min}(r)}^{M_{\rm max}(r)} dM'\ \phi(M') 
\int _0^{\theta _{\rm max}r} dy\ y \Xi (y;r,M,M') 
\label{C2a},\end{equation} 
where $\Xi $, an integrated TPCF, is 
  
\[ 
\Xi(y;r,M,M')
\]\begin{equation}=
\int_{r-\Delta r}^{r+\Delta r}dr' 
\xi\left (\sqrt{y^2+(r-r')^2};r,M,M'\right) 
.\label{Xi}\end{equation}

\subsection{ Further approximations} 

In order to simplify the above calculations, it will be assumed that
the distribution of stars does not 
depend on their luminosity, i.e. that $\xi (y;r,M,M')=\xi (y;r)$. 
This is not completely true as there is a small
dependence on the distribution of stars in 
a cluster according their masses, and the luminosities are dependent on
the masses. A complete calculation taking the luminosity function  into consideration
would be 
of great value. However, the relationship between 
the TPCF and the luminosity function is uncertain, although
the effects of this approximation for the detection of clusters
are expected to be small. 

With this approximation,  and from (\ref{sigmat2}) and (\ref{omegat2}), 
 
\begin{equation} 
\langle\sigma _{\rm t}\rangle=\int _0^{\infty}dr\ \langle N^*\rangle(r) 
\end{equation} 
and
\begin{equation} 
\omega _{\rm t}(\theta)=\frac{1}{\langle\sigma _{\rm t}\rangle^2} \int _0^{\infty}dr\ \langle N^*\rangle^2(r) 
\Xi (r\theta ;r)  
,\label{omegat*}\end{equation} 
where $\langle N^*(r)\rangle$ is the number of stars observed per unit 
solid angle at a distance $r$:
 
\begin{equation} 
\langle N ^*\rangle(r)=r^2\langle\rho \rangle(r) 
\int _{M_{\rm min}(r)}^{M_{\rm max}(r)} dM\ \phi(M) 
.\label{N*}\end{equation} 
 
The variable $\omega _{\rm t}$ can also be expressed as 
 
\begin{equation} 
\omega _{\rm t}(\theta)=\overline{\Xi (\overline{r}\theta )} 
\label{omegaavXi}\end{equation} 
where the averages $\overline{r}$ and $\overline{\Xi }$ are such 
that match (\ref{omegat*}).  
 
Also, from (\ref{C2a}), 
 
\begin{equation} 
C_2=\frac{1}{\langle\sigma _{\rm t}\rangle^2\theta _{\rm max}^2} 
\int _0^{\infty }dr\ \frac{\langle N^* (r)\rangle^2} {r^2} 
\int _0^{\theta _{\rm max}r} dy\ y \Xi (y;r) 
.\label{C2*}\end{equation} 
 
\noindent 
This last equivalence is a way of averaging the function $\xi $. 
 
Hence, high values of $C_2$ indicate that  
there must be high projected clustering in the direction 
of the beam.

\subsection{Patchiness of extinction}
\label{.extincorr}

It is clear that extinction can distort the observed counts, the
amount of the distortion being a matter of controversy. It is  generally
accepted that in the optical wavelengths this influence is very severe,
particularly in regions near to or in the Galactic plane in the inner
Galaxy, where the strong and patchily distributed obscuration makes it
difficult to penetrate deep into the Galaxy. 
The amount of extinction decreases substantially with  increasing
wavelength. Maihara et al. (1978) quoted a value of 0.17
mag kpc$^{-1}$ as  typical for  extinction in the Galactic
plane in the $K$ band, compared with 1.9 mag
kpc$^{-1}$ for the $V$ band (Allen 1973).

This has two important consequences. First, the $K$ band is more
effective at penetrating the interstellar dust. Secondly,  the
observed stellar dis\-tri\-bu\-tion more closely resembles the true
distribution. For the second argument to be true it is necessary 
that the obscuration in the $K$ band should not only be smaller in amount
than in the $V$ band, but also that its patchiness should be less
important.

This rather uniform distribution of the interstellar extinction in $K$
can be inferred from the TMGS histograms in several cuts across the
Galactic plane. Garz\'on et al. (1993, their Fig. 8) compared the
observed stellar distribution in the TMGS and the
GSC in the $V$ band. It is noticeable how uniform the $K$ histograms
are, particularly when  compared with those for the GSC. Except for small
portions highly concentrated in the Galactic plane and more marked in
the central regions, the shape of the high spatial resolution
distribution curves of the TMGS does not exhibit the `noisy' pattern of the GSC
plots, which is certainly due to the presence of strong and patchily distributed
extinction.

Hammersley et al. (1994)   showed similar histograms for
different areas which also have similar shapes. Moreover, a good fit
to a classical ex\-po\-nen\-tial disc can be seen in Fig. 3 of that paper;
this would not be the case if the extinction were important and
non-uniform.

This conclusion can also be reached from the contour maps of the
bulge of the Galaxy of Dwek et al (1995), who showed the residuals of
the DIRBE data after disc subtraction and extinction correction.
Again, the general shape of the maps proves the basic uniformity of
extinction distribution in the near infrared.

We now estimate these effects.
From (\ref{sigmat2}) with 
the change of variable $r=10^{(5+m_{\rm max}-M_{\rm max})/5}$ and 

\begin{equation}
\Phi (M_{\rm max})=\int _{-\infty} 
^{M_{\rm max}} dM\ \phi(M) \approx 
\int _{M_{\rm min}}^{M_{\rm max}} dM\ \phi(M)
,\end{equation}
the local cumulative counts $\sigma _{\rm t}$ follow 
the  expression

\[
\sigma _{\rm t}=\langle\sigma _{\rm t}\rangle _{local}
=200(\ln\ 10)10^{3m_{\rm max}/5}\omega \int_{-\infty }^\infty dM_{\rm max}
\]\begin{equation}\times
 D\left(
10^{(5+m_{\rm max}-M_{\rm max})/5}\right) 10^{-3M_{\rm max}/5}
\Phi (M_{\rm max}),
\end{equation}
ignoring the variation of extinction with the distance.
If we take the density $D$ as  constant, then

\begin{equation}
\sigma _{\rm t}= N(m_{\rm max})\propto 10^{3m_{\rm max}/5}
.\end{equation}
Taking $D$ as  constant
is sufficient for  estimating the  the order of magnitude
of the patchiness due to extinction. 
In any case, the above proportionality 
is followed in the observed cumulative counts but
with a constant value of between 1 and 2 instead of
$3/5$ in the exponent.

An excess of extinction, $\Delta a(\theta)$, due for example to a cloud
at an angular distance $|\vec{\theta } -\vec{\theta _0}|$ with 
respect to a given point $\theta _0$, will cause a reduction 
in the apparent flux of a fraction, $f$, of stars (behind the
cloud), thereby creating the same effect as a  reduction in  
the maximum apparent magnitudes
of these stars by $\Delta a(\theta)$ mag, or, 
if $\Delta a(\theta)$ is 
relatively small, a  reduction in $m_{\rm max}$ by $f\Delta a(\theta)$
mag for all stars. Hence,

\begin{equation}
\sigma _{\rm t}(\theta)\sim \sigma _{\rm t} (\theta _0)10^{-3(a(\theta
)-a(\theta _0))f/5}
\label{sigmada}
.\end{equation}

If it is assumed that the observed flux fluctuations, $\Delta F$, are  
due mainly to extinction variations, with the small-fluctuation
approximation, then  both are related
by

\begin{equation}
\Delta a=-2.5\log \left( 1-\frac{1}{f}\frac{\Delta F}{F} \right)\approx
\frac{5\log _{10}e}{2f}\frac{\Delta F}{F}
\end{equation}
(the factor $f$ appears again here for the same  reasons as above).
So, from  equation (\ref{sigmada}), using the small-fluctuation approximation,

\begin{equation}
\frac{\sigma _{\rm t}(\theta )}
{\sigma _{\rm t} (\theta _0)} \sim 
\frac{3}{2} \frac{F(\theta)}{F(\theta _0)}
\label{omegaext}
.\end{equation}
This means that the angular correlation of star density is about 3/2 times the
angular correlation of the flux.

Averaging the DIRBE $K$ flux (Boggess et al. 1992) 
fluctuations  from the maps with
$2520''$ resolution over
$\mid b\mid \le 3^\circ$ for constant-$l$ strips over the range
$-35^\circ < l< 35^\circ $ (where the effects of extinction are most relevant),
we get root mean squares of

\begin{equation}
0.03< \sigma (\Delta F/F)_{l={\rm const.}, \mid b \mid \le 3^\circ} < 0.23
,\end{equation}
with an average of
\begin{equation}
\overline{\sigma (\Delta F/F)
_{l={\rm const.}, \mid b \mid \le 3^\circ}}= 0.10
.\end{equation}
The oscillations of flux fluctuations are not very high
in the plane, their maximum being $2.3$ times the average.

From equation (\ref{omegaext}), and taking into account that the root mean
square is $\sqrt{\omega (0)}$,

\begin{equation}
\overline{\omega (0)
_{l={\rm const.}, \mid b \mid \le 3^\circ}} \sim 0.015
\end{equation}
for regions of 2520$''$ in size.
In the most unfavourable case, where the extinction is
highest (multiplied by a factor of 2.3$^2$ because the maximum root mean square
is $2.3$ times greater than the average), $\omega (0) \sim 0.08$.

Higher-resolution flux maps are not available in the  $K$-band for the
whole sky so we cannot derive these numbers for smaller scales, but
they are not expected to be much higher since average cloud size
is of the order of degrees (rather higher than 42$'$) 
and the cloud distribution is fairly smooth.
A fractal distribution would increase the contamination but this
may apply only to very cold gas clouds (Pfenninger \& Combes 1994)
which are not the main cause of extinction in the  $K$ band.

We conclude that extinction in $K$
cannot be responsible for correlations $\omega (0)$ greater than
$\sim 0.08$. This is just an estimate, but the order of magnitude should not be
very
different. As will be shown, the results when applied to the TMGS
are above this value (see, for example, Fig. \ref{Fig:a052}), and causes other
than patchy extinction 
must explain this.

\section{ A simple clustering model} 
\label{.simple}

In order to gain an understanding of how the accumulation
parameter ($C_2$) varies with  Galactic position,
a model of stellar distribution is required. 
The model to be adopted in this section is very simple and
consists of a group of spherical star clusters separated by  distances  much
larger than the sizes of the clusters and embedded in a 
Poissonian distribution of field stars with average density 
$\rho _{\rm nc}$ . The density of clusters 
is $\langle n_{\rm cl}\rangle$. To simplify the problem,  a constant star
density, $\rho _{\rm cl}$, is assumed for each  
cluster ($\rho _{\rm nc} < \rho _{\rm cl}$); it is also assumed
that the clusters have the same radius, $R_{\rm cl}$ (i.e. they are filled homogeneous spheres). 
In fact, the density of clusters, their radii and their internal stellar 
densities are different  in different regions of the Galaxy, however there
is insufficient information to construct a more detailed model, and our
main interest here is in how  $C_2$ varies qualitatively.

According to the definition, the TPCF is the average of the product  
of two numbers (for a distance $y$): 
the first is the probability of finding an object in a given position, and the  
second is the number of times that the object counts in a shell of 
radius between $r$ and $r+dr$ exceeds the same counts in a Poissonian  
distribution. The first number, the probability of finding 
an object in a volume $dV$ centered on $\vec{r}$ within a total 
volume $V$,  
is $\rho (\vec{r})dV/ \langle\rho\rangle V$. 
With regard to the second number (Fig. \ref{Fig:twocircles}), there are two cases to be considered: 

\begin{figure} 
\begin{center} 
\mbox{\epsfig{file=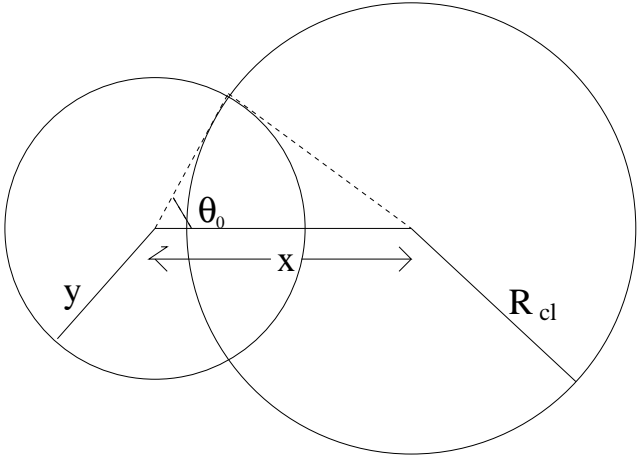,height=6cm,angle=-90}} 
\end{center} 
\caption{ Intersection of two spheres.} 
\label{Fig:twocircles} 
\end{figure}

\begin{enumerate}

\item When $\vec{r}$ is at distance $x<R_{\rm cl}+y$ from the centre of 
a cluster, the second number is the sum of the excess  
(with respect the average, i.e. $\langle\rho\rangle$)
of  objects in the part of the shell that is inside the cluster 
($S_{\rm ss}(x;y,R_{\rm cl})/4\pi y^2 \times (\rho _{\rm cl}-\langle\rho\rangle) 
/\langle\rho\rangle$) minus the deficit of objects in the part of the shell 
outside 
the cluster ($(4\pi y^2 -S_{\rm ss}(x;y,R_{\rm cl}))/4\pi y^2
\times (\langle\rho \rangle-\rho _{\rm nc})/\langle\rho \rangle$).
Here $S_{\rm ss}(x;y,R_{\rm cl})$ is 
the area of a spherical surface of radius $y$ inside another 
sphere of radius $R_{\rm cl}$ whose centre is at distance $x$ from the   
first one.
 
\item When the distance, $x$,  from any cluster is larger than $R_{\rm cl}+y$, 
the second number is the negative quantity $-(\langle\rho\rangle - \rho _{\rm nc})/
\langle\rho\rangle$, i.e. the deficit of objects compared with a Poissonian 
distribution with density $\langle\rho\rangle$. 
 
\end{enumerate} 
 
When $y$ is sufficiently  large  there will be cases in which the shell 
intersects more than one cluster. However,  such large  
values of $y$ are of no interest here, and only those cases in which 
the magnitude of  $y$ is of the same order as that of $R_{\rm cl}$ will be considered. 
 
Thus, the expression for the TPCF is 

{\small 
\[ 
\xi (y)=\langle n _{\rm cl}\rangle 4\pi \int _0^{R_{\rm cl}}dx\ x^2\frac{\rho _{\rm cl}} 
{\langle\rho\rangle}
\]\[ 
\times \left [\frac{S_{\rm ss}(x;y,R_cl)(\rho _{\rm cl}-\langle\rho\rangle )- 
(4\pi y^2-S_{\rm ss}(x;y,R_{\rm cl}))(\langle\rho\rangle -\rho _{\rm nc})}{4\pi y^2\langle \rho \rangle} 
\right ]
\]\[ 
+\langle n _{\rm cl}\rangle 4\pi \int _{R_{\rm cl}}^{R_{\rm cl}+y}dx\ x^2\frac{\rho _{\rm nc}} 
{\langle\rho\rangle} 
\]\[\times 
\left [\frac{S_{\rm ss}(x;y,R_cl)(\rho _{\rm cl}-\langle \rho\rangle )- 
(4\pi y^2-S_{\rm ss}(x;y,R_{\rm cl}))(\langle\rho\rangle -\rho _{\rm nc})}{4\pi y^2\langle\rho\rangle} 
\right ] 
\]\begin{equation} 
-\left [1-\langle n_{\rm cl}\rangle\frac{4}{3}\pi (R_{\rm cl}+y)^3\right ] 
	\frac{\rho _{\rm nc}}{\langle\rho\rangle}\left (\frac{\langle\rho\rangle -\rho _{\rm nc}}{\langle\rho\rangle} 
\right ) 
.\label{ximodel}\end{equation} }
\noindent 
The average density is 
 
\begin{equation} 
\langle\rho\rangle =\frac{4}{3}\pi R_{\rm cl}^3\langle n_{\rm cl}\rangle\rho _{\rm cl}+\left (1- 
\frac{4}{3}\pi R_{\rm cl}^3\langle n_{\rm cl}\rangle \right )\rho _{\rm nc} 
\label{rhoaverage} 
,\end{equation} 
\noindent 
so 
 
\begin{equation} 
\frac{\rho _{\rm nc}}{\langle\rho\rangle}=\frac{1-\frac{\rho _{\rm cl}}{\langle\rho\rangle} 
\frac{4}{3}\pi R_{\rm cl}^3\langle n_{\rm cl}\rangle}{1-\frac{4}{3}\pi R_{\rm cl}^3\langle n_{\rm 
cl}\rangle} 
,\label{rhonc}\end{equation} 
 
\begin{equation} 
\frac{\rho _{\rm cl}-\rho _{\rm nc}}{\langle\rho\rangle}=\frac{\frac{\rho _{\rm cl}}{\langle\rho 
\rangle}-1} 
{1-\frac{4}{3}\pi R_{\rm cl}^3\langle n_{\rm cl}\rangle} 
\end{equation} 
\noindent 
and 
 
\begin{equation} 
\frac{\langle\rho\rangle -\rho _{\rm nc}}{\langle\rho\rangle}=\frac{4}{3}\pi R_{\rm cl}^3\langle n_{\rm cl}\rangle  
\frac{\frac{\rho _{\rm cl}}{\langle \rho \rangle}-1} 
{1-\frac{4}{3}\pi R_{\rm cl}^3\langle n_{\rm cl}\rangle} 
.\end{equation} 
 
Also, from the appendix A1: 
 
{\small  
\[ 
\int _0^{R_{\rm cl}}dx\ x^2S_{\rm ss}(x;y,R_{\rm cl})
\]\begin{equation}=
\left \{ \begin{array}{ll} 
	\frac{4}{3}\pi R_{\rm cl}^3y^2\left (1-\frac{3}{4}\frac{y}{R_{\rm cl}}+  
\frac{1}{16}\left (\frac{y}{R_{\rm cl}}\right )^3\right ),& \mbox{ $y < 2R_{\rm 
cl}$} \\ 
 
	0,& \mbox{ $y \ge 2R_{\rm cl}$} 
\end{array} 
\right \} 
\end{equation} }
\noindent 
and

\[
\int _{R_{\rm cl}}^{R_{\rm cl}+y}dx\ x^2S_{\rm ss}(x;y,R_{\rm cl})
\]\begin{equation}=
\left \{ 
\begin{array}{ll} 
	\frac{4}{3}\pi R_{\rm cl}^3y^2\left (\frac{3}{4}\frac{y}{R_{\rm cl}}- 
\frac{1}{16}\left (\frac{y}{R_{\rm cl}}\right )^3\right ), & \mbox{ $y < 2R_{\rm 
cl}$} \\ 
 
	\frac{4}{3}\pi R_{\rm cl}^3y^2, & \mbox{ $y \ge 2R_{\rm cl}$} \\ 
\end{array} 
\right \} 
.\end{equation} 
 
We insert the last five equalities in (\ref{ximodel}) and, after 
simplifying, this leads to: 

{\tiny

\begin{equation} 
\xi (y)=\left \{ \begin{array}{ll} 
   \langle n_{\rm cl}\rangle\left (\frac{\rho _{\rm cl}-\rho _{\rm nc}}{\langle\rho\rangle}\right )^2 
\frac{4}{3}\pi R_{\rm cl}^3 \\
\times \left (1-\frac{4}{3}\pi R_{\rm cl}^3\langle n_{\rm 
cl}\rangle -\frac{3}{4} 
\frac{y}{R_{\rm cl}}+\frac{1}{16}\left (\frac{y}{R_{\rm cl}}\right )^3\right ), 
& \mbox{$y < 2R_{\rm cl}$} \\ 
 
   -\langle n_{\rm cl}\rangle\left (\frac{\rho _{\rm cl}-\rho _{\rm nc}}{\langle\rho\rangle}\right )^2 
\frac{4}{3}\pi R_{\rm cl}^3\left (\frac{4}{3}\pi R_{\rm cl}^3\langle n_{\rm cl}\rangle \right 
),
& \mbox{$y \ge 2R_{\rm cl}$} \\ 
 \end{array} 
\right \}
\label{ximodel2}\end{equation}   
}

The reader should bear in mind
that the applicable value of $y$  which  is that for 
distances smaller than the minimum distance between two clusters. 
Due to the properties of the TPCF, the 
quantity $\int _{\rm all\  space}dV_y \xi (y)=4\pi \int_0^{\infty} dy\ y^2 
\xi (y)$ should be equal to zero. This is not exactly the case in 
(\ref{ximodel2}), which is not valid for $y$ as large as the typical 
distances among clusters. Nevertheless, it can be seen that 
$\int _{V_y=\langle n_{\rm cl}\rangle ^{-1}}dV_y\xi (y)=0$, i.e. the volume 
in which  (\ref{ximodel2}) can be used  is roughly
$\langle n_{\rm cl}\rangle ^{-1}$. 
 
\begin{figure} 
\begin{center} 
\mbox{\epsfig{file=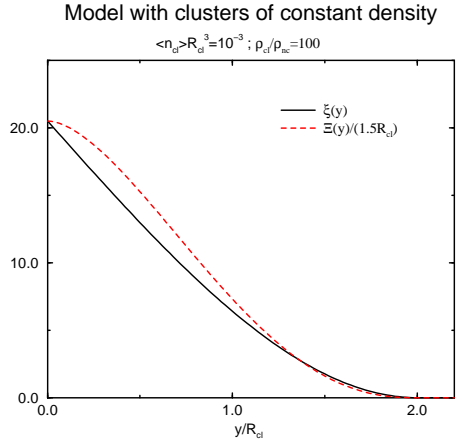,height=6cm}} 
\end{center} 
\caption{ Two-point correlation function, $\xi $, and integrated two-point 
correlation function, $\Xi$, for a simple model. } 
\label{Fig:ximodel} 
\end{figure}

The behaviour of the function can be seen in the Fig. \ref{Fig:ximodel} for  
$\langle n_{\rm cl}\rangle R_{\rm cl}^3=10^{-3}$,
 $\rho _{\rm cl}/\rho _{\rm nc}=100$. 
It becomes constant for $y>2R_{\rm cl}$, and the value in which 
$\xi =0$ is $y=1.90R_{\rm cl}$, which also is very close to $2R_{\rm cl}$. 
In what follows, the term $\frac{4}{3}\pi R_{\rm cl}^3\langle n_{\rm 
cl}\rangle$ will be neglected 
because in practice it is too small (the separation among 
clusters is much larger than $R_{\rm cl}$). 

The next step is the calculation of the integrated TPCF ($\Xi$)
using  (\ref{Xi}) and 
including (\ref{ximodel2}). In this case, $\Delta r$ is the first value 
that follows $\xi (\sqrt{y^2+(r-r')^2})=0$, which is $\Delta r= 
\sqrt{4R_{\rm cl}^2-y^2}$. Hence, 
 
{\tiny
\[ 
\Xi \left(t=\frac{y}{2R_{\rm cl}}\right)= 
\int_{r-\sqrt{4R_{\rm cl}^2-y^2}}^{r+\sqrt{4R_{\rm cl}^2-y^2}}dr' 
\xi(\sqrt{y^2+(r-r')^2})
\]\[=
\langle n_{\rm cl}\rangle
\left (\frac{\rho _{\rm cl}-\rho _{\rm nc}}{\langle\rho\rangle}\right )^2 
\frac{4}{3}\pi R_{\rm cl}^3 4R_{\rm cl}
\]\begin{equation} \times
\left[\sqrt{1-t^2}\left(\frac{3}{8}+ 
\frac{3}{16}t^2 \right)+ 
\ln \left(\frac{\sqrt{1-t^2}+1}{t} 
\right )\left (\frac{-3}{4}t^2+\frac{3}{16}t^4 \right) 
\right ].
\end{equation} }

 Fig. \ref{Fig:ximodel} shows that the behaviour of the TPCF and the integrated TCPF   ($\xi $ and $\Xi $) are not very different. 
 
Now, using (\ref{C2a}), the observed  value $C_2$ will be calculated. The theoretical lower limit of the integral 
should be zero, but in practice values under 
a $y_{\rm min}=r\theta _{\rm min}$ cannot be observed, due to the resolution 
$\theta _{\rm min}$ of our detector ($r$ is the distance of 
the zone of clusters). 
 
\[ 
\frac{1}{r^2\theta _{\rm max}^2}\int _{r\theta _{\rm min}}^{r\theta _{\rm max}} 
dy y \Xi (y) 
\]\[ 
=\left(\frac{2R_{\rm cl}}{r\theta _{\rm max}}\right)^2 
\langle n_{\rm cl}\rangle \left (\frac{\rho _{\rm cl}-\rho _{\rm nc}}{\langle\rho\rangle}\right )^2 
\]\begin{equation} \times
\frac{4}{3}\pi R_{\rm cl}^3\left[\frac{1}{6}\ R_{\rm 
cl}\left(F\left(\frac{r\theta_{\rm min}} 
{2R_{\rm cl}}\right)-F\left(\frac{r\theta_{\rm max}} 
{2R_{\rm cl}}\right)\right)\right]  
,\label{deltaNNmodel}\end{equation} 
\noindent 
where $F(t_0)$ is  
 
{\tiny
\[
F(t_0)=24\int_{t_0}^1dt\ t 
\]\begin{equation} \times 
\left[\sqrt{1-t^2}\left(\frac{3}{8}+ 
\frac{3}{16}t^2 \right)+ 
\ln \left(\frac{\sqrt{1-t^2}+1}{t} 
\right )\left (\frac{-3}{4}t^2+\frac{3}{16}t^4 \right) 
\right ] 
\end{equation} } 
($F(0)=1$, $F(1)=0$).

The observational value of $C_2$ would be an average of  
this quantity with weight $\langle N^*\rangle ^2(r)$, according to 
(\ref{C2*}), and with the characteristic parameters of the cluster 
depending on $r$, i.e. $\langle n_{\rm cl}\rangle (r)$ and  
$\left( (\rho _{\rm cl}-\rho _{\rm nc})/\langle\rho\rangle\right )^2(r)$.
 
\subsection{The contribution of a single shell of 
clusters} 
\label{singleshell}

In the case where there  are clusters only in the shell between $r$ 
and $r+\delta r$, 
with $\delta r/r\ll 1$, and where the contribution to $C_2$ 
is  given only for this range of $r$ (when in the other ranges of  $r$
the contribution is nil, or  when the other shell with clusters 
has a negligible $\langle N^*\rangle(r)$), then
 
\begin{equation} 
r\theta _{\rm max} \approx 2R_{\rm cl}, 
\label{thetamaxmodel}\end{equation} 
because the TPCF ($\xi$) in (\ref{ximodel2}) is zero for this value 
(as $\frac{4}{3}\pi R_{\rm cl}^3\langle n_{\rm cl}\rangle$ is much smaller than unity). Then, introducing  
(\ref{thetamaxmodel}) in (\ref{deltaNNmodel}) and feeding the result 
into (\ref{C2*}), we obtain 
 
\[
C_2=\frac{\langle N^*\rangle^2(r)\delta r}{\langle\sigma _{\rm t}\rangle ^2}\langle n_{\rm cl}\rangle 
\left (\frac{\rho _{\rm cl}-\rho _{\rm nc}}{\langle\rho\rangle}\right )^2 
\frac{4}{3}\pi R_{\rm cl}^3 
\]\begin{equation}\times
\left[\frac{1}{6}\ R_{\rm 
cl}F\left(\frac{\theta_{\rm min}} 
{\theta_{\rm max}}\right)\right] 
\label{C2clus1} 
.\end{equation} 
 
The factor $F$ reduces the value of $C_2$ when $\theta _{\rm max}$ is only  
few times greater than $\theta _{\rm min}$. Since $\theta _{\rm 
max}=2R_{\rm cl}/r$, the more distant the clusters the smaller the value obtained for $C_2$. 
$\theta _{\rm max}$ is normally enough large compared with 
$\theta _{\rm min}$ for $F$ to be considered as always close to unity. 
The values that  $F$ takes are shown in Table \ref{Tab:F}. 

\begin{table}[htb] 
\begin{center} 
\caption{ Correction factor $F$ due to the finite resolution of the 
detector.} 
\begin{tabular}{|c|c|} 
\\ \hline 
$x$ & $F(x)$ \\ \hline 
0 & 1.000000 \\ 
0.1 & 0.956356  \\ 
0.2 & 0.838143  \\  
0.3 & 0.671930  \\  
0.4 & 0.487970  \\  
0.5 & 0.314330  \\  
0.6 & 0.172488  \\ 
0.7 & 0.074429  \\ 
0.8 & 0.021008  \\  
0.9 & 0.002144  \\  
$\ge 1$ & 0  \\ \hline 
\label{Tab:F} 
\end{tabular} 
\end{center} 
\end{table} 

As can be seen, when $x$ is greater than $\approx 1/4$, i.e. when 
the distance of the cluster is greater than 
$\sim R_{\rm cl}/2\theta _{\rm min}$, the effect of the factor $F$ 
begins to be noticeable. 
 
\subsection{Clusters distributed throughout the Galaxy} 

Suppose that there are clusters distributed throughout the Galaxy, i.e. 
that there are  clusters at all distances along the line of sight
in any direction. The TPACF
($\omega _{\rm t}$), which 
is related to the TPCF ($\xi $) through (\ref{omegat*}), 
is computed numerically. The function $\xi (y)$ 
is obtained with (\ref{ximodel2}), where it is assumed that the size and star density in the cluster is 
the same  in all clusters, and that also  the density of clusters, $\langle n_{\rm cl}\rangle$,
is proportional to the mean density of stars: 
 
\begin{equation} 
\langle n_{\rm cl}\rangle =C\langle\rho\rangle . 
\end{equation} 
 
The density of stars ($\rho _{\rm nc}$) is inferred 
from the relationship (\ref{rhonc}).
$\langle N^*\rangle(r)$ is calculated using (\ref{N*}), where $\langle\rho\rangle$
and $\phi (M)$ are calculated for a model Galaxy with two
components: a disc and  a bulge. The disc is taken  from the
Wainscoat et al. (1992) model and the bulge model described in
L\'opez-Corredoira et al. (1997). 
The extinction law given in Wainscoat et al. (1992) is also used.

The explicit dependence of $C_2$ on $l$ and $b$ is calculated from 
the $\omega$ values and a $\theta _{\rm max}$ that is derived with 
the approximation $\theta _{\rm max} \approx 2R_{\rm cl}/\overline{r}$,
where 
 
\begin{equation} 
\overline{r}=\frac{\int_0^\infty dr\ r N^*(r)}{\int_0^\infty dr\ N^*(r)} 
.\end{equation} 
 $\theta _{\rm max}$ is calculated in this way because the first 
zero of the function $\omega $  cannot be derived. 
In the  previous approximations, $\omega $  was always positive and
 negative values  were  neglected . 
 
\begin{figure} 
\begin{center} 
\mbox{\epsfig{file=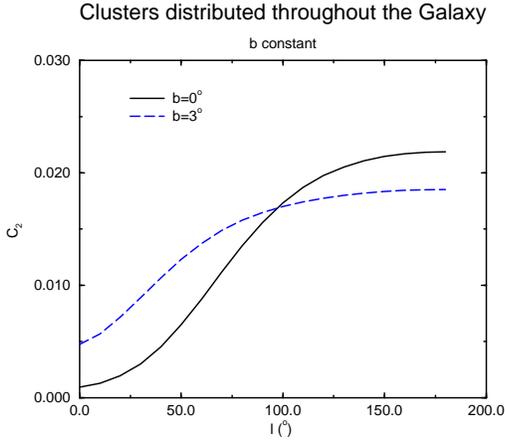,height=6cm}} 
\end{center}
\caption{ $C_2$ at $b=0^\circ $, $b=3^\circ $ and $b=10^\circ$
according to a distribution of simple clusters over all the Galaxy. } 
\label{Fig:m2b0} 
\end{figure}

\begin{figure} 
\begin{center} 
\mbox{\epsfig{file=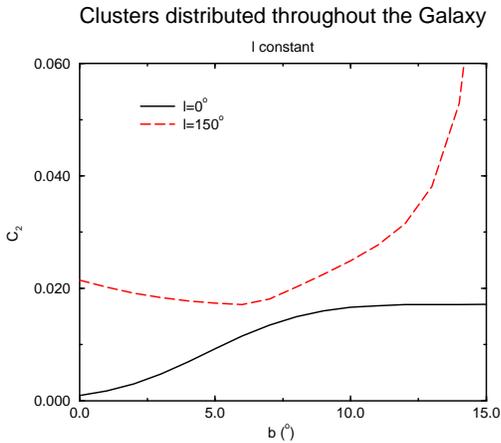,height=6cm}} 
\end{center} 
\caption{ $C_2$ for different Galactic latitudes for $l=0^\circ $
and $l=150^\circ $ 
according to a distribution of simple clusters over all the Galaxy. } 
\label{Fig:m2l0} 
\end{figure}

When these operations are carried out for the values $R_{\rm cl}=1$ pc, 
$\rho _{\rm cl}=500\ pc^{-3}$ and $C=10^{-5}$,  
the results shown in Figures \ref{Fig:m2b0} and \ref{Fig:m2l0} are obtained. 

In Figures \ref{Fig:m2b0} and \ref{Fig:m2l0} it can be seen how the
 correlation  increases with distance from the Galactic plane 
(increasing $|b|$) as well as with distance from the centre 
of the Galaxy (increasing 
$|l|$). An explanation of this can be seen  directly in the expression 
(\ref{ximodel2}), which is proportional to  
$\langle n_{\rm cl}\rangle \left ((\rho _{\rm cl}-\rho _{\rm nc})/\langle\rho \rangle\right )^2$, 
where $\rho _{\rm cl}$ is a much larger constant than 
$\rho _{\rm nc}$ and $\langle n_{\rm cl}\rangle$ is proportional to 
$\langle\rho\rangle$. It can be seen that $\xi $ is greater for smaller values 
of $\langle\rho\rangle$, i.e. it is greater away from 
the Galactic plane and far away from the bulge. Since the 
angular correlation function is an average of all the correlation 
functions in the line of sight, it produces the above results.  
 
This is merely a hypothetical example, and no meaning should be given to the actual 
values obtained since the parameters are invented. However, the qualitative dependence of
 $\omega$ on $l$ and $b$ is significant. This result is even more general 
than the particular case of the proportionality $\langle n_{\rm cl}\rangle =C\langle \rho \rangle$. 
An increase in $\omega$ for increasing $|l|$ or $|b|$ is given when  
$\langle n_{\rm cl}\rangle\left (({\rho _{\rm cl}-\rho _{\rm nc})/\langle\rho \rangle}\right )^2$ 
increases in a typical region in the line of sight. Even a dependence such as 
$\langle n_{\rm cl}\rangle =C\langle\rho\rangle ^\alpha$ with $\alpha <2$ is acceptable to 
allow this kind of dependence. Another significant prediction 
is that in the anticentre  
region the dependence with $l$ is very smooth and  nearly constant.

\section{The causes of the clustering} 
 
The existence of clustering indicates that the formation of the stellar 
components 
of a cluster share the same time and place of birth, and that all the stars 
that are observed in the cluster 
once belonged to same originating cloud, which explains why they occupy a 
neighbouring position in space. The alternative hypothesis of an initial 
Poissonian 
distribution of stars which collapsed to form the cluster 
is not feasible because 
cluster's gravity would have been too weak  to restrain the velocities of the 
stars, which would have ``evaporated'' from the primitive cluster. 
The fact is, then, that clusters originate from earlier clusters 
that were 
formed from a cloud, or several clouds in the same region. 
 
It is assumed in the TPCAF, equation (\ref{omegat2}), that 
only $\xi$, and not the extinction through the relationship between 
apparent magnitudes and absolute magnitudes, is dependent on $\theta $.  
As discussed in section \ref{.extincorr}, the contribution from cloud irregularity
is irrelevant.
  
\subsection{ Relationship with the evolution time} 
\label{.clustime}

The relationship of the accumulative parameter with
the evolution time comes from the rate of evaporation of stars.

When the effects of dust clouds are ignored, all contributions 
to $C_2$ come from $\xi $, i.e.   clusters that happen to be in the line of sight. 
The relationship with the evolution time is through $\rho _{\rm cl}$ since 
stars escape from the cluster over time. 
According to Chandrasekhar (1942), the rate of escape of stars is 
 
\begin{equation} 
\frac{\Delta \rho _{\rm cl}(t)}{\rho _{\rm cl}} \approx -Q\frac{\Delta t}{T_{\rm E}},
\end{equation} 
where $Q$ and $T_{\rm E}$ are  constants that depend on the characteristics of 
the cluster. $Q$ is the rate of stars which can  escape and 
$T_{\rm E}$ is the average time that takes these stars to leave the cluster.  
$Q$ is the fraction of stars with velocities greater than escape 
velocity. Chandrasekhar (1942) calculates a value of $Q=0.0074$ 
for a relaxed cluster.  
$T_{\rm E}$ (in Gyr) is the relaxation time that is, for an average cluster 
with $N$ stars, radius $R$ (in pc)  
and average stellar mass $\overline{m}$ (in $M_\odot$)
 
\begin{equation} 
T_{\rm E}=8.8\times 10^{-4} \sqrt{NR^3/\overline{m}}\frac{1}{\log _{10} N-0.45}. 
\end{equation} 
 
Insofar as $C_2$ is proportional to $\left( (\rho _{\rm cl}-\rho _{\rm 
nc}) 
/\langle\rho\rangle\right)^2$ through the proportionality dependence with $\xi $ in 
(\ref{ximodel2}) and assuming $\langle n_{\rm cl}\rangle$ and $R_{\rm cl}$ as constant and 
$\rho _{\rm cl}\gg\langle\rho\rangle$, then $C_2$ may be approximated as 
 
\begin{equation} 
C_2(t) =C_2(t=0)  
e^{-\frac{Qt}{2T_{\rm E}}} 
,\end{equation} 
or, expressed differently, 
 
\begin{equation} 
C_2(t)= K_1\times e^{-t/K_2} 
,\end{equation} 
where $K_1$ and $K_2$ are positive constants. 
 
This is the theory for simple cases but some cases are more
 complicated. Certain 
other effects 
are not negligible, such as  dynamical friction 
(see Chandrasekhar 1943a,b,c) or  Galactic rotation and
gravitational tides (Wielen \& Fuchs 1988), 
both of which produce different values of 
these constants.  
Spitzer (1958) points out that most open clusters should be destroyed by   
interactions with molecular clouds on time-scales of few hundred million   
years, meaning that few very old open clusters are known to have  
survived to the current epoch. However,  $N$-body simulations (Terlevich 1987)  
predict that 
only by encounters with the most massive molecular clouds would the cluster  
be disrupted. 
 
Apart from theoretical considerations, exponential decrease is indicated by   
other authors from observational data (for example, Lyng\aa \ 1987a). 
Janes \& Phelps (1994) fit a relationship between the age and cluster  
abundance in the solar neighbourhood which follows a sum of two exponentials.   
If it is assumed that there is a constant creation of clusters, and that the 
death of a cluster corresponds to a low value of $\xi $,   this 
would then imply that $C_2$ is dependent on the sum of two decaying exponentials, 
although with  so many effects it is difficult to determine which is 
the correct dependence. What is clear, however, is that high values 
of $C_2$ indicate the existence of young clusters. 
 
\section{ Measurement of correlations in the TMGS and other surveys}

\subsection{ Measurement of the TPACF} 
 
The following discussion concerns the measurement of the TPACF 
derived from a rectangular field image of angular size $a\times b$ in a direction containing $N$ 
stars of known coordinates and magnitudes  (between 
$m_{\rm min}$ and $m_{\rm max}$). 
 
One method of determining correlation functions in a distribution 
of objects,   
discussed by Rivolo (1986), is to use the following estimator for 
$N$ points:

\begin{equation} 
\langle\rho \rho\rangle (r)=\frac{\langle\rho\rangle}{N}\sum_{i=1}^N \frac{M_i(r)}{V_i(r)}, 
\end{equation} 
where $M_i (r)$ is the number of particles lying in a shell of thickness  
$\delta r$ from the $i$th particle, and $V_i (r)$ is the volume of the shell. 
The same applies to $\langle\sigma \sigma\rangle$ but with 
areas instead of volumes: 
 
\begin{equation} 
\langle\sigma \sigma\rangle (\theta)=\frac{\langle\sigma\rangle}{N}\sum_{i=1}^N  
\frac{M_i(\theta)}{\Omega _i(\theta)}.
\label{Rivoloang}\end{equation}  
 
This expression must be corrected for edge effects, i.e. 
 some stars are lost in the calculation of $M_i(\theta)$ when a star  
$i$ is at distance less than $\theta $ from an edge of the rectangular image. 
In the quantity $\sum_{i=1}^N M_i(\theta )$,  
the excess probability is measured of finding a star at an angular distance 
$\theta $ from other stars in a ring of thickness
$\Delta \theta$ with surface area $\Omega (\theta)$ and this 
should 
be proportional to $\Omega (\theta)$ for a Poissonian distribution. The excess 
probability is reflected when there is an excess of stars inside  
the ring $\Omega (\theta)$. The loss of stars due to edge effects  
is cause by part of the ring falling outside the area of rectangle. 
 
To solve the edge-effect problem it is necessary to calculate how many stars 
are lost beyond the edges with respect to a non-edge case.
Only a fraction $F_{\rm BE}$ is measured 
for stars separated by an angular distance $\theta $,
and  each $M_i(\theta)$ must be divided by $F_{\rm BE}(\theta )$. 
The calculation of $F_{\rm BE}$ 
is given in  Appendix A2 for a rectangle. This is represented by
 
\begin{equation} 
F_{\rm BE}(\theta )= 1 - \frac{2}{\pi}\left(1+\frac{a}{b}\right)\frac{\theta} 
{a} + \left(\frac{9}{\pi}+\frac{\pi }{2}-4\right) \frac{1}{a\ b} \theta ^2 
.\end{equation} 
 When this is applied to (\ref{omega}), it gives 
 
\[ 
\omega_{\rm t}(\theta)=\frac{a\ b\sum_{i=1}^NM_i(\theta )}{N^22\pi\theta d\theta 
\left[1 - \frac{2}{\pi}\left(1+\frac{a}{b}\right)\frac{\theta} 
{a}+\left(\frac{9}{\pi}+\frac{\pi }{2}-4\right) \frac{1}{a\ b} \theta 
^2\right]}
\]\begin{equation}
-1 
.\label{measureomega}\end{equation} 
 
With this simple algorithm, the angular distances of the stars with respect to one another (once their coordinates are known) and 
the angular correlation for a rectangular field 
of stars are obtained. 
 
The error in the TPACF,  as  for the 
TPCF, is derived by Betancort-Rijo (1991). In the limit 
of small $\Delta \theta$ (the interval for calculating the different 
values of $\omega (\theta)$)  the error expression leads to 
 
\[ 
S(\omega_{\rm t} (\theta ))=\frac{1+\omega (\theta)}{(\sum_{i=1}^NM_i(\theta )) 
^{1/2}}\] 
\begin{equation} 
=\frac{a\ b\sqrt{\sum_{i=1}^NM_i(\theta )}} 
{N^22\pi\theta 
d\theta\left[1-\frac{2}{\pi}\left(1+\frac{a}{b}\right)\frac{\theta} 
{a}+\left(\frac{9}{\pi}+\frac{\pi }{2}-4\right) \frac{1}{a\ b} \theta ^2\right]}. 
\end{equation} 
 
After this, the different integrals containing 
$\omega _t$, and  their errors, can be calculated with standard numerical algorithms.

\subsection{ Some examples with known clusters in visible}

The TPACF, as a statistical tool for inferring correlation,
is applicable to any survey.  Some test examples are given below
in order to determine  how good the method is at  finding regions of 
the sky with clusters, both where there known to be one
or two open clusters and  where no clusters have been 
identified. Some
open clusters were randomly selected from Messier catalogue and are listed in Table 
\ref{Tab:clusvis}. These are six regions with one Messier open cluster, 
two regions with two Messier open clusters and two regions with none.
Stars down to magnitude 12 in $V$ were selected from the GSC.

\begin{table*}[htb] 
\begin{center} 
\caption{Regions selected for deriving correlations.} 
\begin{tabular}{|c|c|c|c|c|c|}\\ \hline   
Objects & $\alpha _{1950.0}$(h) & $\delta _{1950.0}$(deg) & Size($'$)
& $\theta _{\rm max}$($'$) & $C_2$ \\ \hline
M29 & 20.368 & 38.367 &  7 & 9 & 0.356$\pm $0.059 \\
M34 &  2.647  & 42.567  & 35.2 & 19 & 0.118$\pm $0.012 \\
M35 &  6.097  & 24.350  & 28 & 37 & 0.190$\pm $0.005 \\
M39 & 21.507  & 48.217  & 32 & 25 &  0.049$\pm $0.004 \\
M45 &  3.733  & 23.967  & 110 & 143 & 0.043$\pm $0.001 \\
M52 & 23.367  & 61.317  & 18 & 7 & 0.329$\pm $0.063 \\
M36 \& M38 &  5.478  & 34.950 &  12 \& 21 & 21 & 0.130$\pm $0.014 \\
M46 \& M47 &  7.615  & -14.533 &  27 \& 30 & 21 & 0.141$\pm $0.008 \\
None &  5.478  & 36.950  & - & 13 & 0.034$\pm $0.016 \\
None &  7.615 &  -12.583  & - & 5 & -0.002$\pm $0.050 \\ \hline
\label{Tab:clusvis} 
\end{tabular}  
\end{center} 
\end{table*} 

A square field
three times larger than the catalogued size of the largest 
cluster was selected and
$\omega _{\rm t}$ (hereafter called simply $\omega $) was derived from 
equations (\ref{Rivoloang}) and (\ref{omega}).
The value of $\theta _{\rm max}$ is derived as the angle whose $\omega $  
is zero within the error; for larger values of $\theta $, 
 $\omega $ is more or less equally positive and negative. This criterion is 
not precise when the errors are large but it gives an acceptable estimate. 
$C_2$ is obtained for each region from equation (\ref{C2}). 

\begin{figure} 
\begin{center} 
\mbox{\epsfig{file=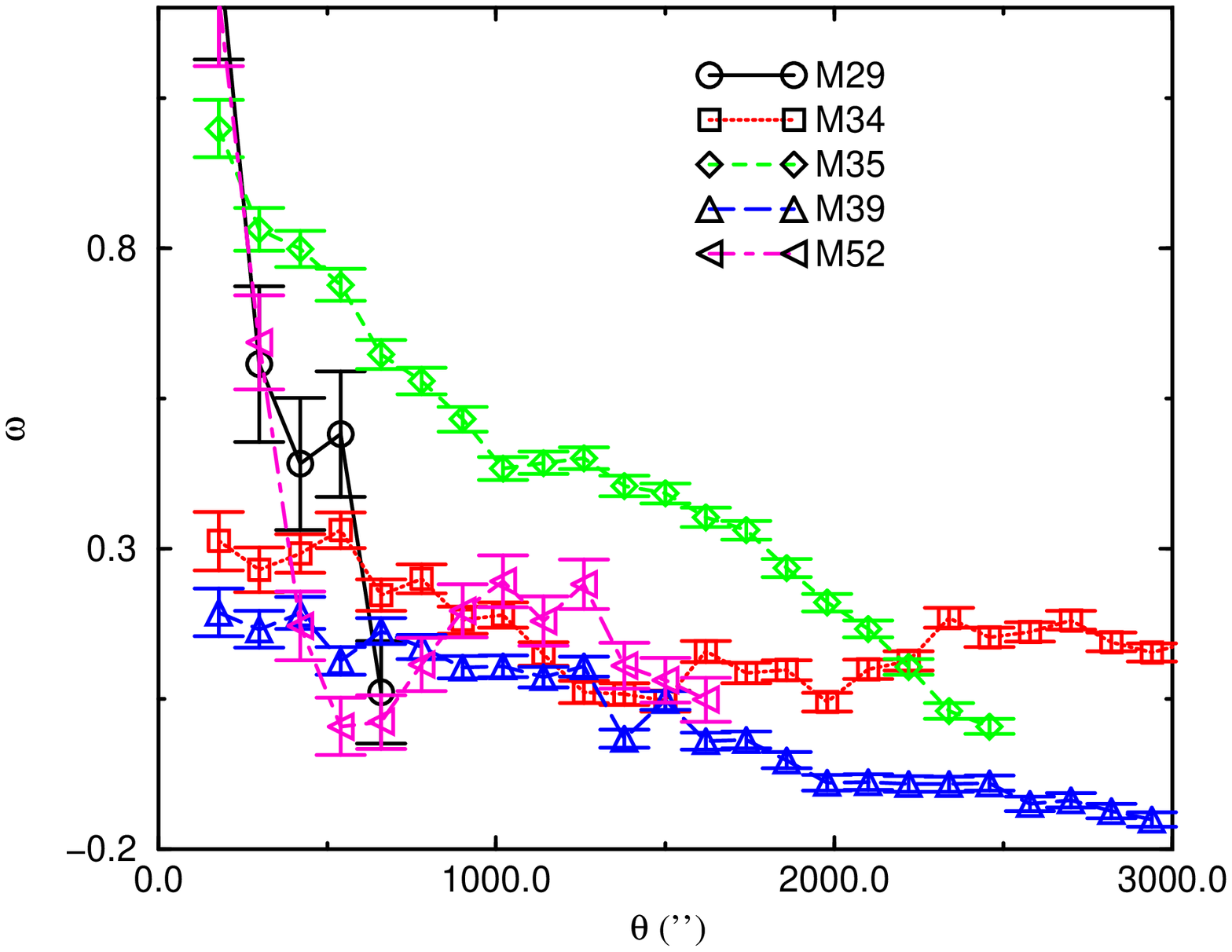,height=6cm}} 

a)
\end{center} 
\caption{ $\omega $ at some regions with visible magnitudes.} 
\label{Fig:angcorrvis} 
\end{figure}

 Figures \ref{Fig:angcorrvis}{\it a-c}
show the TPACF, and Table \ref{Tab:clusvis}
lists $\theta _{\rm max}$ and $C_2$. The correlations are positive to within the errors for scales shorter  than the size of
the clusters. Since the correlations have been calculated with stars
down to magnitude 12,  many of the stars do not belong to the cluster.
Because of this, the correlation is not excessively high, although it is high 
enough to
distinguish it from the cases with no clusters (Fig.
\ref{Fig:angcorrvis}{\it c}), which represents
two random cases in regions without clusters two degrees to the north of
the two fields with two clusters each. 
In the first field without clusters there is a small correlation but it is insignificant to within $3.5\sigma $. 

\begin{figure} 
\begin{center} 
\mbox{\epsfig{file=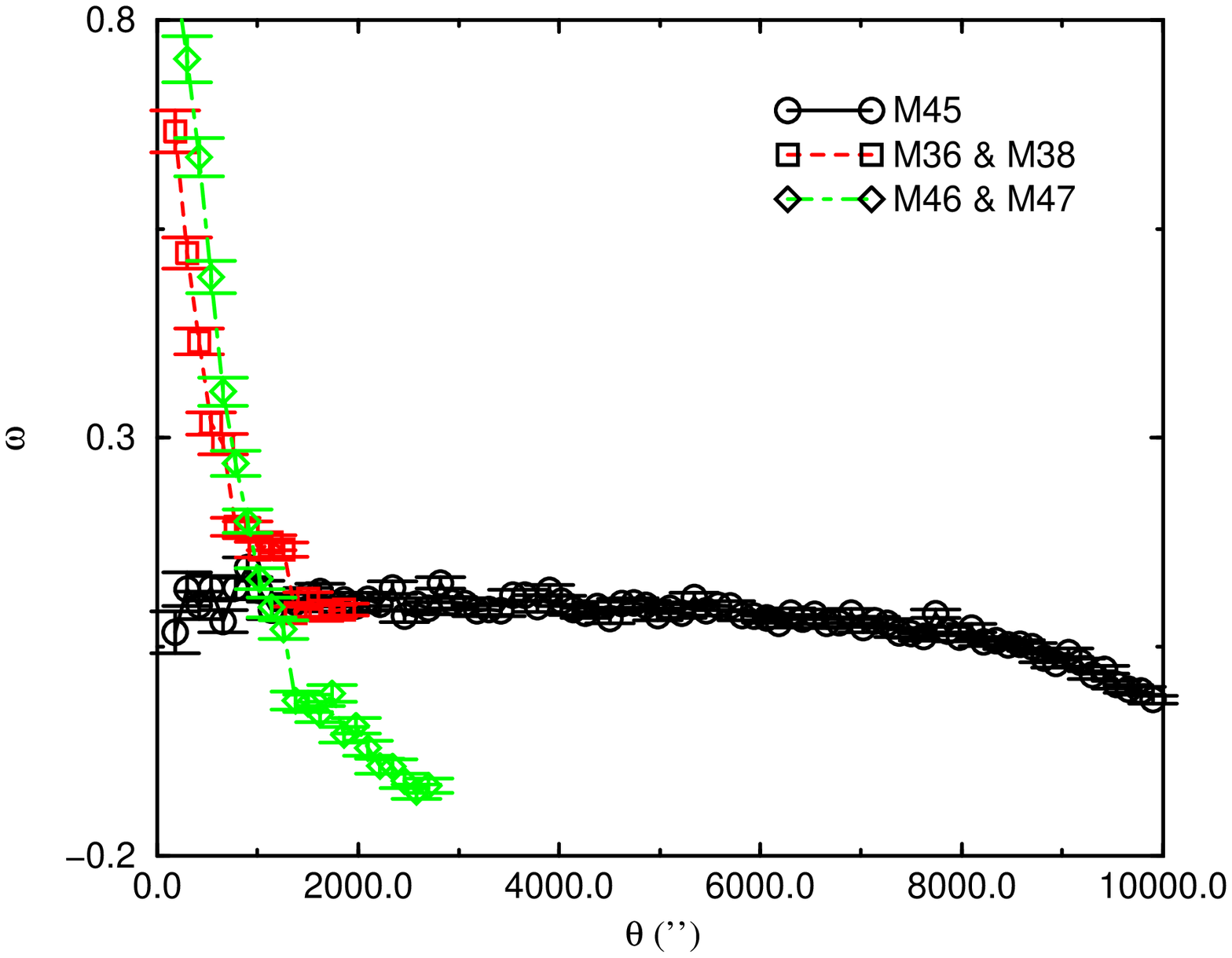,height=6cm}} 

Fig. \ref{Fig:angcorrvis} b)
\end{center} 
%\caption{ $\omega $ at some regions with visible magnitudes.} 
%\label{Fig:angcorrvis2} 
\end{figure}

\begin{figure} 
\begin{center} 
\mbox{\epsfig{file=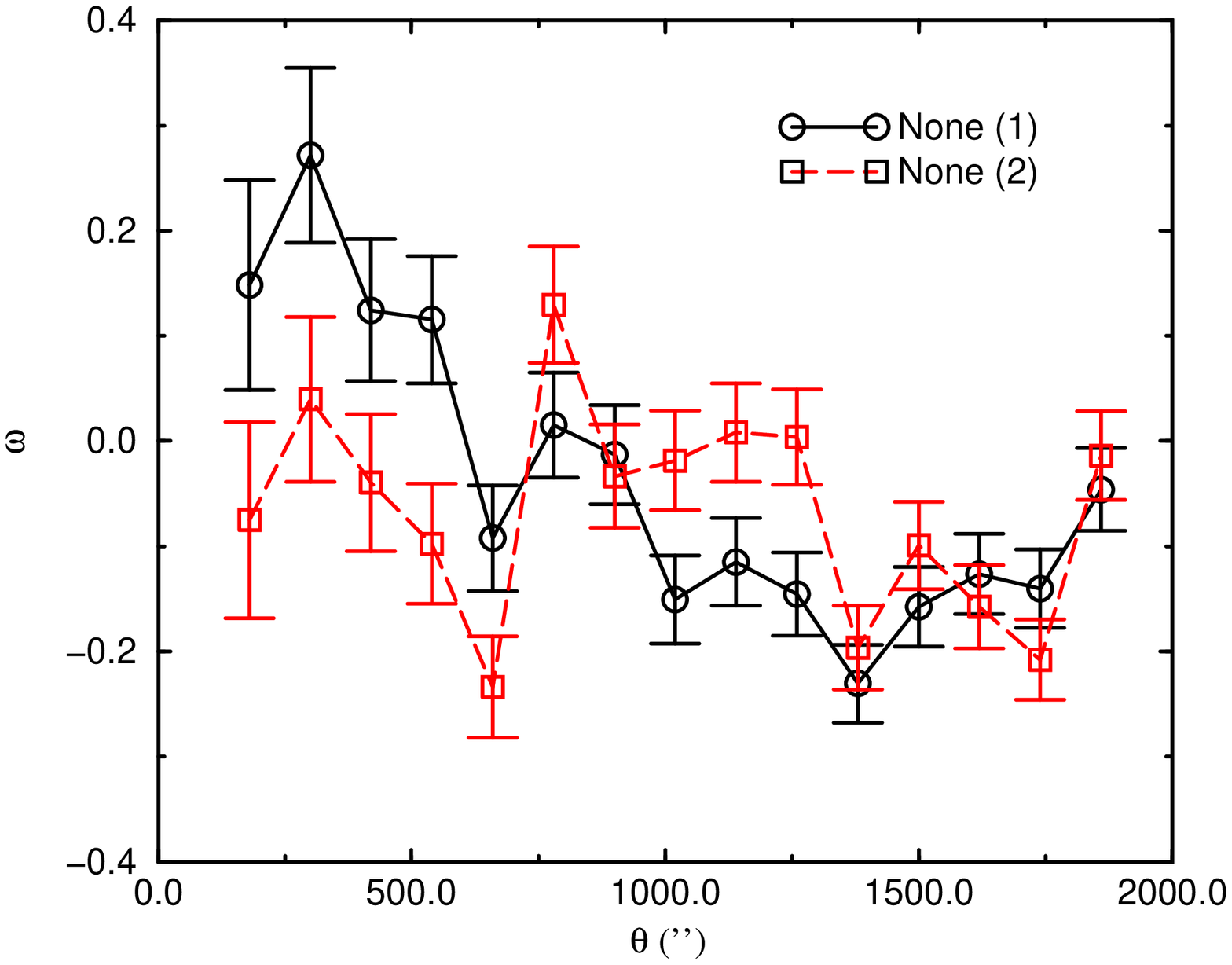,height=6cm}} 

Fig. \ref{Fig:angcorrvis} c)
\end{center} 
%\caption{ $\omega $ at some regions with visible magnitudes.} 
%\label{Fig:angcorrvis3} 
\end{figure}  

\begin{figure} 
\begin{center} 
\mbox{\epsfig{file=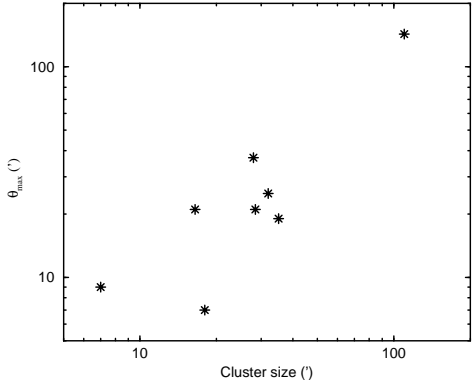,height=6cm}} 
\end{center} 
\caption{$\theta _{\rm max}$, the predicted size of the average cluster
according to the simple model used in this paper versus the catalogued
size of the cluster in this region (in case of two clusters, 
the average size of both is plotted).} 
\label{Fig:sizes} 
\end{figure}

Hence, the method does indeed detect clustering where there are known to be
clusters but not where there is believed to be none. One cluster would be enough if its effect were not too
attenuated by foreground and background stars in the chosen range of
magnitudes.  Moreover, the predictions of the  sizes of the clusters is quite acceptable in comparison
with the catalogued sizes (see Table \ref{Tab:clusvis} and Figure
\ref{Fig:sizes}).

\subsection{Peculiarities of TMGS data in calculating  the 
TPACF} 
 
Due to the way in TMGS data is obtained, the following 
considerations must be taken into account when applying the method
and in the examination of the results obtained. 
 
\begin{itemize} 
 
\item The method of assigning declinations to the stars in the TMGS will give 
extra angular correlations in the angles which are multiples of the
 angular size 
of the detector (17 arcsec) or multiples of the quarter diameter. 
Declinations are assigned with discrete values, which are
separated by distance multiples of a quarter of the detector size ($\sim 4''$).
However, the correlation using $\theta$ greater than 4 or 5
arcsec is not significantly  affected by this characteristic. 
The range of $\theta $
to be used in this paper is from $5$ to $250$ arcsec; moreover,
$C_2$ is averaged over a wide range of $\theta $, much higher than
4 arcsec, so this effect is negligible for this particular case.  
However, for  the correlation of angles smaller than $4$ arcsec this 
effect could be important.

\item The strips do not cover the whole sky; neither do they  completely  
cover 100\% of the area of the rectangles that we will use to calculate the 
angular correlation function. The TMGS (Garz\'on et al. 1993) was carried out 
by means of drift scanning with strips 
of constant declination, and in some cases there are small gaps between adjacent
strips 
which make the sky coverage within the squares in the region of interest  
incomplete. 
To correct for this effect, we must multiply 
$(1+\omega _{\rm t}(\theta ))$ by the fraction of area  covered in the rectangle with 
regard 
the expression (\ref{measureomega}), 
assuming that the positions in the rectangle which are not covered is random. 
The fraction of  area covered in the squares that we use is high 
(greater than $80$ \%) so the measure of 
the error is good enough because it is only affected by a factor of between  
$\sqrt{0.80}$ and unity. 
 
\end{itemize}  

\subsection {Application to TMGS} 
\label{.ejemplocorrTMGS}

As an example, two cases will now be applied  to the TMGS.

\begin{figure} 
\begin{center} 
\mbox{\epsfig{file=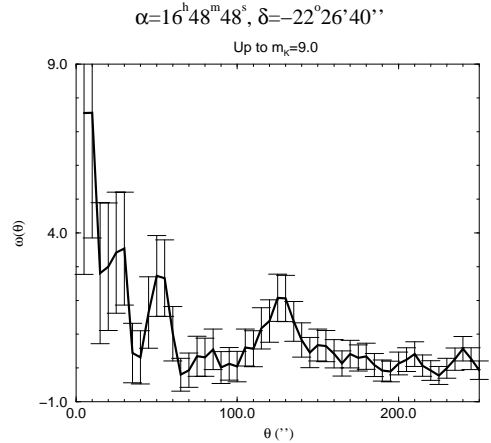,height=6cm}} 
\end{center} 
\caption{ $\omega $ at $l=-1^\circ58'15''$, $b=14^\circ12'13''$. } 
\label{Fig:a052} 
\end{figure}

From the TMGS data a rectangular region of the sky was selected  centred on 
$\alpha =16^{\rm h}48^{\rm m}48^{\rm s}$, $\delta=-22^\circ 26'40''$ (J2000.0, corresponding
to Galactic coordinates $l=-1^\circ 58'15''$, $b=14^\circ12'13''$) and sides 
$0.462^\circ$ and $1.626^\circ $, respectively, for the directions of $\alpha $ and  
$\delta $. In this field, the calculation of the TPACF was carried out for  stars 
down to 9.0 mag in  $K$.  
From (\ref{Rivoloang}) and (\ref{omega}), we obtain $\omega $ up to
$\theta =250''$, as presented in Figure \ref{Fig:a052}. 
An upper limit of $250''$ for the value of 
$\theta _{\rm max}$ is considered.   
The parameters used in the measure are $\theta _{\rm min}=5''$ and  
$\Delta \theta=5''$. 

The result after calculating  $\omega $,  using  (\ref{C2}) is 
  
\[ 
\theta _{\rm max}=62.5 ''\pm 2.5'' 
\]
and\[ 
C_2=1.21 \pm 0.22.
\] 
 This is an example of a field where some correlation is found ($C_2$ is 
non-zero at the  5.5-$\sigma $ level). Clearly, there is  
clustering rather than a Poissonian distribution. 
 
As a second example, a rectangular region was chosen 
centered on 
$\alpha =18^{\rm h}50^{\rm m}24^{\rm s}$, $\delta=-5^\circ 14'7''$ (J2000.0;  $l=28^\circ 10'18''$, $b=-2^\circ 8'40''$)   with sides 
$0.498^\circ$ and $0.289^\circ$, respectively, for the direction ($\alpha $,$\delta $). 
Calculation of the angular correlation function was also made for the stars 
down to 9.0  mag in $K$ for this field.  
 
\begin{figure} 
\begin{center} 
\mbox{\epsfig{file=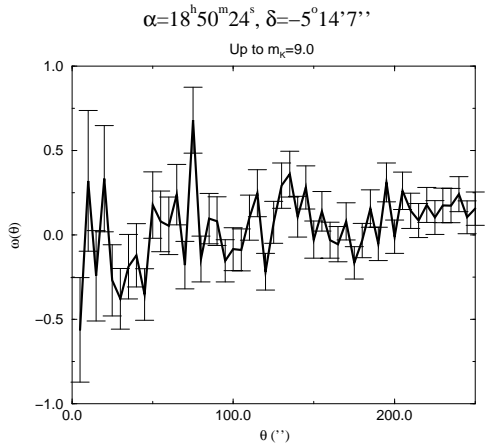,height=6cm}} 
\end{center} 
\caption{ $\omega $ at $l=28^\circ 10'18''$, $b=-2^\circ 8'40''$.} 
\label{Fig:a252} 
\end{figure}

Using (\ref{Rivoloang}) and (\ref{omega}), and with the same parameters as for 
the previous case, the  $\omega$ shown in Fig. \ref{Fig:a252} is obtained. This
gives finally:
 
\[ 
\theta _{\rm max}=22.5'' \pm 2.5'' 
\]
and\[ 
C_2=3.0\times 10^{-2} \pm 14.6 \times 10^{-2}.
\] 

In this case the correlation is very weak ($C_2$ is $\sim 40$ times smaller  
than in the previous case), and there is a difference of only 
$0.2$ $\sigma$ from zero, which, within the errors, implies  that there 
is no correlation or clustering among the stars from this 
field. Figure \ref{Fig:a252} shows that 
$\omega $ is almost zero 
for every value of $\theta $ within the error bar. The value of $\theta _{\rm 
max}$ is  
meaningless in this case, because $\omega $ has such a low value 
that the error in 
the search for the first zero of the function is large. When there is little
correlation in the field $\theta _{\rm max}$ will be small 
because the algorithm that eliminates the zeros due to fluctuations 
does not work well when $\omega $ is much smaller than its error. 
Also, in this case
$\sigma (C_2)$, the error in $C_2$, will be larger than, or of same order as, 
$C_2$.  To overcome this problem those data with large $\sigma (C_2)$
will be separated  and only  those with $C_2>3.5\times \sigma (C_2)$
will be considered as having confirmed correlations.

\subsection{ Which clusters can be detected with this procedure?}
\label{.limitecorr}

In the following application, the TPACF will be measured for stars down to $K=
 9.0$ mag and for a maximum angle of $250''$. From Appendix A2, this
requires a minimum strip width of  $\sim 500''$.
Larger clusters would be detected with larger angles; however, $250''$ is 
almost the 
largest angle that can be used if the same analytical criteria are to 
be applied to all
ten TMGS strips that we will use.

A typical open cluster has an average size of $5$ pc (Janes \& Phelps 1994),
so $500''$ will be sufficient to detect it at $\sim 2$ kpc.  Therefore, 
mostly distant clusters will be detected. However, the TMGS is dominated
by late K and M giants. Therefore, the majority of the stars detected in 
the magnitude range to be used are significantly further away than 2~kpc.
It is estimated that  $\sim 10'$ is the maximum size of the clusters which
will affect  $C_2$, although this is difficult to calculate
accurately, and the contribution of the largest clusters is not
nil but decreases as the cluster increases in size.

The exact calculation 
of the minimum size is also
difficult to analyse. The TMGS survey does not detect the individual sources in  globular clusters since the stars are too close to be separated and the
whole cluster will appear as an extended source. Similarly, if the open 
cluster is very
small  there is excessive
 overcrowding, or confusion, of sources, a few of them 
 contributing very little to the parameter $C_2$. The separation between 
the stars in a cluster needs to be more than twice the diameter
of the detector for its components to be detected,
i.e. a minimum  of $30''$ to $35''$.

In conclusion, it is expected to detect distant clusters with angular diameters
of between 0.5 and 10 arcmin. Solar
neighbourhood clusters, such as M45,  will  provide only a small 
contribution to the
quantity $C_2$, so they will not be detected. Hence the analysis will focus on the
large-scale distribution of clusters in the Galaxy.

\section{Correlations as a function of Galactic coordinates}

When the  procedure is used 
in  other regions, the behaviour of $C_2$ as a function  
of $l$ and $b$ can be determined. 
 
Calculations of $\omega (\theta )$ were carried out on several strips  
with constant declination and various sub-strips (Table \ref{Tab:zonesdelta}). 
The value of  $l$ quoted  is that at which the strip intersects the Galactic plane ($b=0^\circ$).  
The range of Galactic latitude is $|b|<15^\circ$ for $l<35^\circ$ and
$|b|<5^\circ$ for $l>35^\circ$.

\begin{table}[htb] 
\begin{center} 
\caption{ TMGS regions (with constant declination) used in this 
paper.} 
\begin{tabular}{|c|c|c|c|}\\ \hline   
Strip & $\delta $ & $l$ at $b=0^\circ$  & Width of strip \\ 
&(deg)&(deg)&(deg)\\ \hline 
1 & $-29.7$ & $-0.9$ & $2.51$ \\ 
2 & $-22.4$ & $7.5$ & $1.63$ \\ 
3 & $-15.6$ & $15.4$ & $0.78$ \\  
4 & $-5.2$ & $27.1$ & $0.29$ \\  
5 & $-1.6$ & $31.2$ & $1.50$ \\  
6 & $3.5$ & $36.9$ & $0.13$ \\ 
7 & $22.0$ & $57.8$ & $0.30$ \\ 
8 & $32.5$ & $70.1$ & $0.35$ \\ 
9 & $39.1$ & $167.8$ & $0.35$ \\  
10 & $30.4$ & $178.2$ & $0.13$ \\ \hline
\label{Tab:zonesdelta} 
\end{tabular}  
\end{center} 
\end{table} 

\begin{figure}
\begin{center}
\mbox{\epsfig{file=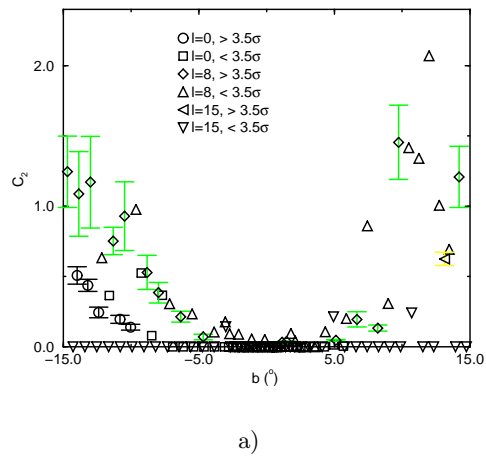,height=6cm}}

a)
\end{center}
\caption{$C_2$ at a) $\delta=-30$, $-22$ and $-16$;
b) $\delta=-5$, $-1$ and $4$ ; c) $\delta=22$, $33$, $39$ (in the anticenter
direction)
and $30$ (in the anticenter direction).
Points with $C_2>3.5\sigma
(C_2)$ are plotted with bar error.}
\label{Fig:corr}
\end{figure} 
 
\begin{figure}
\begin{center}
\mbox{\epsfig{file=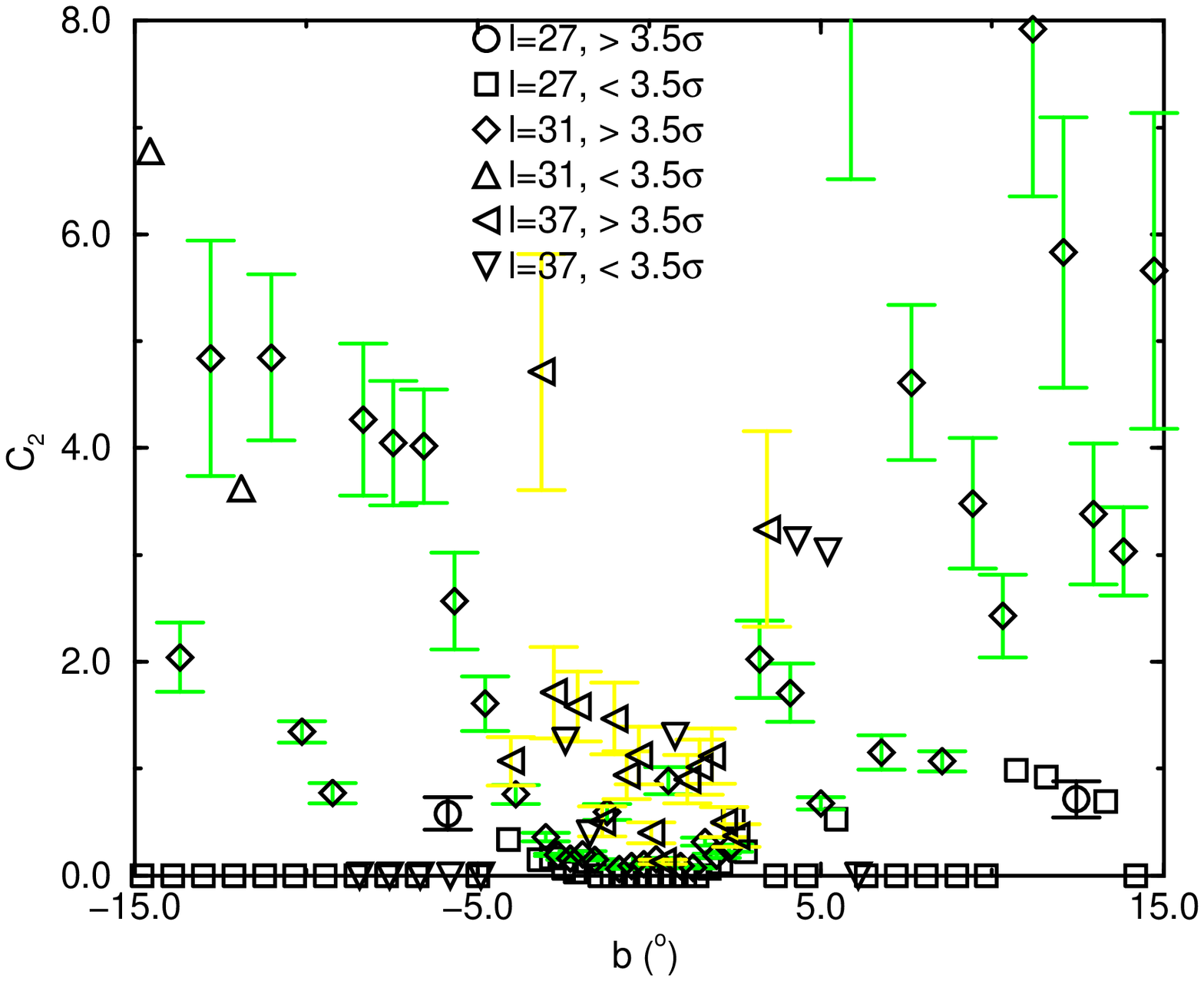,height=6cm}}

Fig. \ref{Fig:corr} b)
\end{center}
\end{figure}
 
\begin{figure}
\begin{center}
\mbox{\epsfig{file=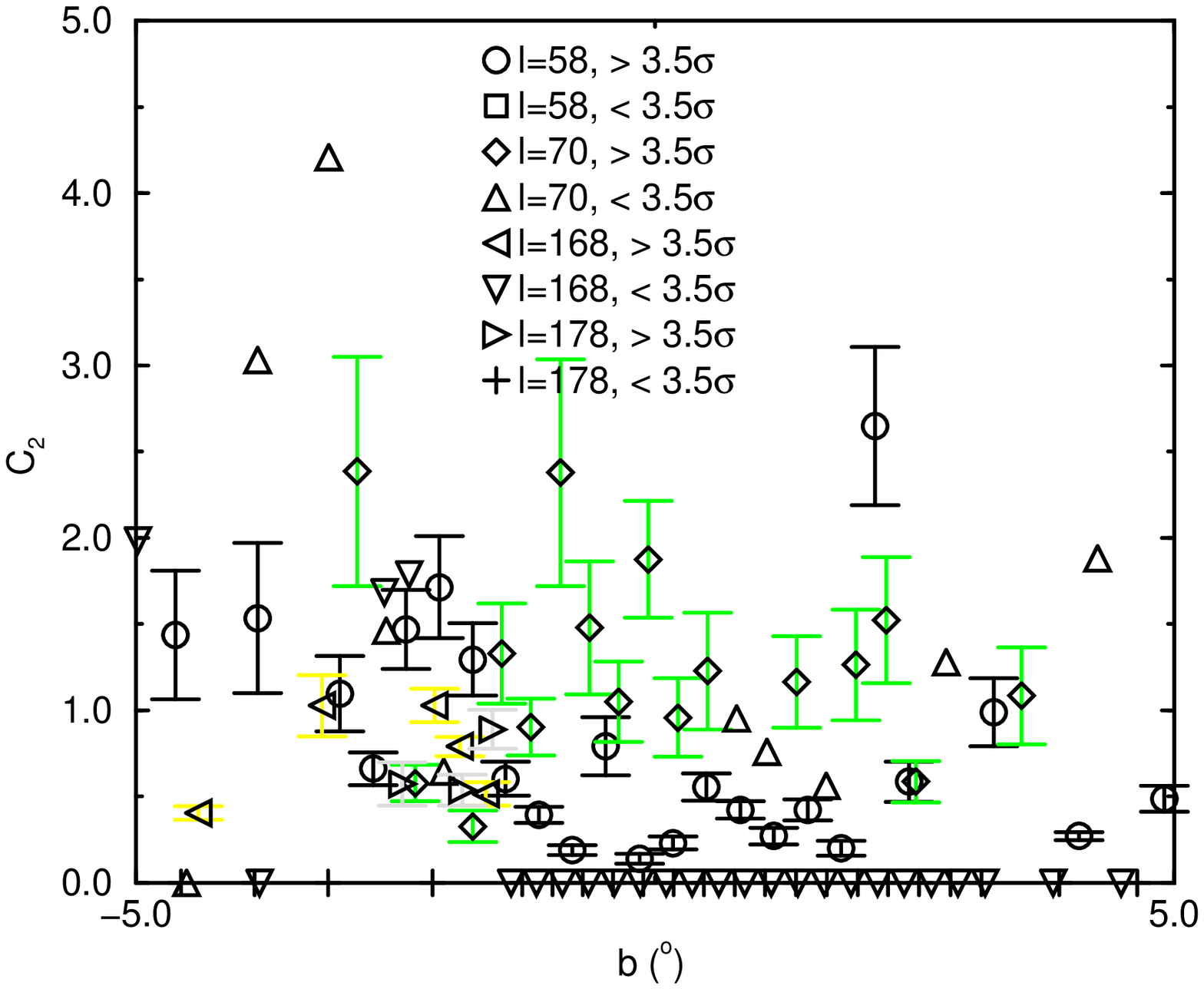,height=6cm}}

Fig. \ref{Fig:corr} c)
\end{center}

\end{figure}

The results are plotted in the Figs. \ref{Fig:corr}, 
 those for which $C_2 > 3.5$ $\sigma (C_2)$ being separated from the 
others. Not all the strips have the same number of stars, so $\sigma (C_2)$  
is different for  different regions. The best strips, with smallest errors,  
are those which cross  
the plane at $l=-1^\circ$, $l=8^\circ$ and $l=31^\circ$, i.e. strips 1, 2 and 5 respectively. The main  features observed are: 

\begin{figure} 
\begin{center} 
\mbox{\epsfig{file=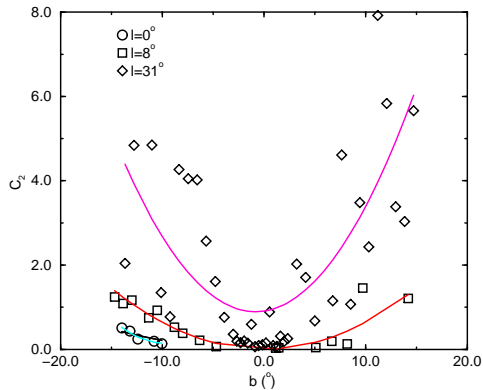,height=6cm}} 
\end{center} 
\caption{ Fit of $C_2$ of strips 1, 2 and 5 (those that are greater than  
$3.5\ \sigma (C_2)$) to respective parabollic functions: $2.2\times 10^{-3}b^2$,
$6.4\times 10^{-3}b^2$ and $20\times 10^{-3}b^2+0.036b+1.0$ (solid lines).} 
\label{Fig:fitsq} 
\end{figure} 

\begin{enumerate} 
 
\item There is general a $C_2$ dependence on Galactic latitude in the
disc for $l< 90^\circ $ (i.e.  strips 1-8 )
with some exceptions pointed out in (iv). This relationship is especially remarkable in strips 
1, 2 and 5. The function $C_2(b)$ is approximately parabolic.
Figure~\ref{Fig:fitsq} shows the following fits to the best data  
(those with $C_2>3.5\ \sigma (C_2)$) of these strips: 
 
\begin{equation} 
C_2=2.2\times 10^{-3}b^2 \ \ {\rm for}\ l=0^\circ 
,\label{C2_l0}\end{equation} 
 
\begin{equation} 
C_2=6.4\times 10^{-3}b^2 \ \ {\rm for}\ l=8^\circ 
\label{C2_l8}\end{equation} 
and

\begin{equation} 
C_2=20\times 10^{-3}b^2+0.036b+1.00 \ \ {\rm for}\ l=31^\circ 
.\label{C2_l15}\end{equation}

\item 
Outside the bulge,
there is a  general  increase in $C_2$ with $l$, as seen in equations
(\ref{C2_l0}), (\ref{C2_l8}) and (\ref{C2_l15}).
When the data with $C_2 > 3.5$ $\sigma (C_2)$ are averaged  
between $b=-3^\circ $ and $b=3^\circ $, 
and also for $3^\circ <|b|<5^\circ $, there is a  dependence on $l$ as shown in   
Table \ref{Tab:depend_l}.

\item
 In the inner bulge region, with $l<15^\circ$ and $b<5^\circ$,   
the correlations are almost zero (Fig. \ref{Fig:corr} a)).
When  the relative correlation differences  at $l\sim 30^\circ$
and the inner bulge are compared with those predicted by the  simple
model (Figs. \ref{Fig:m2l0}, \ref{Fig:m2b0}) a correlation deficit 
 can be seen for the bulge.
 
\item 
Three regions where there is an excess of correlation with 
respect to both the $l$ and $b$ dependence occur at
 at $l=31^\circ$,
$l=37^\circ$ (Fig. \ref{Fig:corr}b) and $l=70^\circ$ 
(Fig. \ref{Fig:corr}c), i.e. strips 5, 6 and 8.  
 
\item 
The anticentre region gives a correlation similar to that of the intermediate 
Galactic longitude ($50^\circ <l<100^\circ $) region  in the plane,
or even lower (see Table \ref{Tab:depend_l}). 
The value of $C_2$  does not increase with $l$, or it does so very smoothly  
outside the intermediate $l$ region. The simple 
model prediction is not 
very accurate but, comparing with the results in Figs. \ref{Fig:m2b0} 
and \ref{Fig:m2l0}, it can be 
seen that $C_2$ is larger at $l \sim 150^\circ$ than  
at $l\sim 40^\circ$. This significant departure cannot be explained without including an  extra  component  in the  model.

\end{enumerate} 

\begin{table}[htb] 
\begin{center} 
\caption{The $C_2$ dependence on $l$.} 
\begin{tabular}{cccc}\\ \hline   
Strip & $l$ at $b=0^\circ$ & $\overline{C_2}$ in $|b|<3^\circ $ 
& $\overline{C_2}$ in $3^\circ <|b|<5^\circ $ \\ 
&(deg)&&\\ \hline 
1 & $-1 $ & -- & --  \\
2 & $8 $ & $3.5\times 10^{-2}$ & $6.9\times 10^{-2}$ \\ 
3 & $15 $ & -- & -- \\ 
4 & $27 $ & -- & -- \\ 
5 & $31 $ & $0.38$ & $1.29$ \\ 
6 & $37 $ & $0.90$ & $3.01$ \\ 
7 & $58 $ & $0.74$ & $0.97$ \\ 
8 & $70 $ & $1.26$ & $1.08$ \\ 
9 & $168 $ & $0.78$ & $0.71$ \\  
10 & $178 $ & $0.67$ & -- \\ \hline
\label{Tab:depend_l}  
\end{tabular}  
\end{center} 
\end{table}

\subsection{Causes of the dependence on Galactic coordinates}

\begin{enumerate} 
 
\item {\it The $b$-dependence:} 
This is predicted by the simple model created above. 
As seen in Fig. \ref{Fig:m2l0},  $\omega $ should increase  with $|b|$   
because $\langle n_{\rm cl}\rangle \left( (\rho _{\rm cl}-\rho _{\rm nc})/\langle\rho \rangle\right) 
^2$  
in expression (\ref{ximodel2}) increases with $|b|$.  
  
\item {\it The $l$-dependence:} 
Again, the  $l$-dependence can be explained  by the model. 

\item {\it The bulge:}
When $C_2$ values for bulge regions are compared
with those for
 other regions where there is only a disc component,
  the observational data give a lower relative correlation than is prediced by the model (Fig. \ref{Fig:m2b0}), i.e. there is less correlation in the bulge than expected. 
This is consistent with the bulge being older than the disc (see  Section   \ref{.clustime}). 
The bulge is known to have a different population 
of stars from  the disc (Frogel 1988; L\'opez Corredoira
et al. 1997), and these are expected to be an older population (Rich 1993), 
so the lack of correlation would be expected. 
Moreover, as pointed out earlier, the central region will have patchy 
extinction which will tend to increase the correlation. Hence, the 
 $C_2$ values generated  by clusters should be even lower than
the observed value indicating even fewer young
clusters in this region.

\item {\it  Excess correlation in  some zones in the plane:}
Feinstein (1995) points out that very young clusters are tracers of the spiral arms. As has been noted young clusters provide a significant contribution to 
$C_2$. However, spiral arms are not included in the simple model, 
which  contains only  the  disc and bulge; 
hence where an arm is crossed it is to be expected that 
the correlation will be higher than predicted. Whereas the distribution 
of old clusters varies smoothly with Galactic position, the
young clusters have a far more irregular distribution. 

Of the four areas which cross the plane between $l=31^\circ $ and $l=70^\circ $ 
three show a significant correlation  excess. 
The $l=31^\circ$ region is almost  tangential to the Scutum arm and also
runs through the   Sagittarius arm.
The  $l=37^\circ $ region will also cut the Sagittarius arm. 
The excess correlation at  $l=70^\circ $ can  be attribute  
to the star formation region in the Perseus arm.
The far-infrared source G69C in this  
region has been attributed to  star formation regions by Kutner (1987). 
Strip 7 ($l=58^\circ $) is an exception which could possibly be due to this line of sight missing a significant star formation region as it crosses the arm.

Using the simple model described previously with one shell of clusters
(Section~\ref{singleshell}) but only allowing  clusters 
in the arms, 
the degree of clustering in the arm can be estimated. With the use 
of (\ref{C2clus1}) 
a relationship between various cluster parameters can be obtained 
(for example, at $l\approx 70^\circ $, $b\approx 0^\circ $, $C_2$ is 
$1.88 \pm 0.34$). It is assumed that  
the greater part  of the contribution to $C_2$ 
is due to the arm clusters, and that, as estimated by Cohen et al. (1997, in preparation) $\sim 30$\%  of the stars
observed in that 
region are from the arm (with 
$\Delta r\approx 1$ kpc in the line of sight), whereas the rest belongs 
to the disc. As $\theta _{\rm max}$ is 
$46.5''$ in this region, the equivalent cluster diameter, from 
(\ref{thetamaxmodel}), is $2.0$ pc (when the distance is $r=8.8$ kpc; 
Georgelin \& Georgelin 1976). This is  smaller than the   
average value 
of $5$ pc given by Janes \& Phelps (1994), although this could possibly be
 due to the fact that  
only the core of the cluster is seen by the TMGS
(the typical size of the core of a cluster is 1 or 2 pc; Leonard 1988), 
in which there are bright 
and more massive stars; alternatively,  a possible difference from the solar  
neighbourhood could also be the explanation. Nevertheless, 
this is only a rough estimate and no further conclusions should be 
drawn from this number. The order of magnitude should, however, be 
correct. 
 
With these data it possible to estimate the density of stars within the clusters. Using  (\ref{C2clus1}) gives
 
\begin{equation} 
\langle n_{\rm cl}\rangle \approx 3\times 10^4 \left(\frac{\rho _{\rm cl}-\rho _{\rm nc}} 
{\langle\rho\rangle}\right)^{-2} 
,\end{equation} 
or, making use of (\ref{rhoaverage}) with $\frac{4}{3}\pi R_{\rm cl}^3 
\langle n_{\rm cl}\rangle\ll 1$ and $\rho _{\rm nc} \ll \rho _{\rm cl}$,  
 
\begin{equation} 
\langle n_{\rm cl}\rangle \approx 3\times 10^4 \left[\frac{\frac{4}{3}\pi R_{\rm cl}^3 
\langle n_{\rm cl}\rangle\rho _{\rm cl}+\rho _{\rm nc}}{\rho _{\rm cl}}\right]^2 
.\end{equation} 
 
When $\langle n_{\rm cl}\rangle$ is determined with last equation, the
following condition is then derived 
for solving the second-degree equation with real numbers: 
 
\begin{equation} 
\rho _{\rm nc} < 2 \times 10^{-6} \rho _{\rm cl} 
\label{l70ronc} 
\end{equation} 
and 
 
\begin{equation} 
5\times 10^{-7}\ {\rm pc}^{-3} <{\rm \langle}n_{\rm cl}{\rm \rangle}< 2 \times 10^{-6}\ 
{\rm pc}^{-3},  
\label{l70ncl} 
\end{equation} 
which is equivalent to saying that most of the stars within the arm in this line of sight are forming clusters. 
Indeed, with such a low $\rho _{\rm nc}$ (because of (\ref{l70ronc})), 
by (\ref{rhoaverage}) 
and (\ref{l70ncl}) gives 
 
\begin{equation} 
10^5  \langle\rho\rangle\ \ <\ \ \rho _{\rm cl}\ \ <\ \ 5\times 10^5 \langle\rho \rangle 
,\end{equation} 
which, with $\langle\rho\rangle\sim 1.4\times 10^{-3}\ {\rm pc}^{-3}$ (From the Galaxy model  used in Wainscoat et al. 1992), gives
 
\begin{equation} 
140\ {\rm pc}^{-3} < \rho _{\rm cl} < 700\ {\rm pc}^{-3}    
,\end{equation} 
i.e. between $500$ and $3000$ stars per cluster, a quite reasonable number
(Friel 1995 talks about a typical mass of young clusters
of few thousand solar masses).

The same case is repeated at $l=37^\circ$ and at $l=31^\circ$. When the line of
sight cuts an arm, the correlation is greatly increased. If the arm
contribution to the number of stars is low, the
 correlation will be diluted,  although its contribution may be not
totally negligible. It also depends on the density of the other
components for that direction; the disc, for example, dilutes the cor\-re\-la\-tion. 

Hammersley et al. (1994) and Garz\'on et al. (1997) suggest that
there should be an excess 
of bright stars in the region at $l=27^\circ $, $b=0^\circ $ 
which might be due to the interaction of a bar with 
the disc, giving rise to a star formation region. 
Concerning the deficit of correlation measured
for this star formation region, it could be concluded that, 
if the star formation were sufficiently large--much greater, say,
 than the size of the rectangle used from sampling--then this would  explain  the  non-detection of clustering in this region.

\item {\it Deviation from a simple model of $l$ dependence in the anticentre:} 
Towards the anticentre ($210^\circ > l> 150^\circ $) the 
 correlation is significantly less than that predicted by the simple model. 
This implies that the number of  clusters is small  and, in particular, 
that very young clusters are rare in this direction. This is in agreement with 
the results from   visible observations of clusters (Payne-Gaposchkin 1979).
Janes \& Phelps (1994) argue that  there is a lack of old 
clusters in the inner disc, since they would be destroyed by molecular clouds 
(see Section~\ref{.clustime}), but that there will a relatively large number 
of young clusters. However, the ISM density falls off with distance from
the Galactic centre, 
so in the outer disc there will be significantly less star formation and 
hence far fewer young clusters. The existence of a gradient in open cluster 
age has been commented on   by Lyng\aa \ (1980) and Van den Bergh \& McClure 
(1980),  and an  explanation was attempted by Wielen (1985).
As has been noted previously, young clusters
contribute significantly to $C_2$, so the lack of young clusters 
in the anticentre region would lead to a reduction in the amount of correlation.

Within a few degrees of the plane the arms have a significant influence on 
the amount of correlation for the longitude range   $30^\circ <l<90^\circ $.
One possible reason for the apparent deficit in $C_2$ towards the anticentre 
could be the excess due to the arms in the comparison regions. In order 
to  discount this possibility, a comparison of the in-plane anticentre 
region can be made with an off-plane region at $l=31^\circ$. The model 
predicts that the ratio $C_2(l$$=$$31^\circ,$$5^\circ$$<$$|$$b$$|$$<$$15^\circ)$$/$$C_2(l$$=$$168^\circ,$$|$$b$$|$$<$$5^\circ)
$  should be less than unity, however, the measurements give a value of 4 or 5.
This gives further support to the hypothesis that there are fewer  
young clusters than expected towards the anticentre.

 A further possible reason for the lack of correlation in the anticentre is that
there could be a significant decrease in the total 
numbers of clusters in the anticentre rather then an increase of age
with Galactocentric distance. However, observations in the solar  
neighbourhood (Lyng\aa \ 1980; Van den Bergh \& McClure 1980; Janes \& Phelps 1994) 
support the hypothesis of increasing age of the clusters with  Galactocentric distance.

Also, clusters in the anticentre have a greater angular size
because they are nearer but this is taken into account in the model and
should not cause the deficit of correlation.

When the analysis presented here is applied to other large-area infrared observations  it may contribute to our understandanding of 
this dependence on cluster age on Galactocentric distance. 
The TMGS data clearly shows the presence of young clusters in the inner Galaxy and consequently 
a decrease in $C_2$ in the anticentre direction. However,
An accurate quantification
is not possible because of 
 arm contamination in most parts of the regions, for which
complete information is unavailable.

\end{enumerate} 
 
\section{ Conclusions} 
 
A technique is developed for  searching for clustering in stellar  surveys 
using  correlation functions. The mathematical 
tools are useful for any field of stars and 
can be applied to any survey, 
especially those at carried out at infrared wavelengths, which
permit a study of the distribution of stars throughout almost the entire 
Galaxy. The DENIS (Epchtein 1997) or 2MASS (Skrutskie et al. 1997)
 surveys will be ideal for this technique as the increased numbers of stars will reduce the errors. It is even possible, 
with a large number of stars in the survey, to apply the technique for
 different ranges of apparent magnitude. Studying the 
 clustering of stars at different 
apparent magnitudes is equivalent to do  studying in three dimensions 
($l$, $b$ and the average distance $\overline{r}$ which is associated with 
the treated range of magnitudes).

A simple model has been developed. This model could be improved by introducing 
a density dependence as a function of the distance from the centre of the 
cluster, perhaps a power-law dependence. 

In this paper the method has been applied to the TMGS. 
Is has been shown that a simple model in which old open clusters trace 
the whole Galaxy 
with a density of clusters proportional to the density of stars agrees 
quite well with the data. An exception to the general agreement  are  
specific regions in the plane where the higher-than-expected clustering can be
a\-ttri\-bu\-ted  to star formation in the spiral arms. 
A second departure from the simple model  is the reduced $C_2$ 
in  the outer disc and in the bulge due to a lack of young  clusters.
 
In one of the regions with an excess, at $l=70^\circ $ in 
the plane, the approximate limits for  
the cluster density and the density of stars inside the cluster are derived. 
These are, respectively,  
$5\times 10^{-7}\ {\rm pc}^{-3} <{\rm \langle}n_{\rm cl}{\rm \rangle}< 2 \times 10^{-6}\ 
{\rm pc}^{-3}$ 
and $140\ {\rm pc}^{-3} < \rho _{\rm cl} < 700\ {\rm pc}^{-3}$.  
There is, however, 
a lower-than-expected  correlation  at $l=27^\circ $, $b=0^\circ $. 
There is believed to be a huge star formation  region in this direction
and the lack of correlation could be 
due to the star formation region being far larger than the sample area.
  
As has been pointed out by Friel (1995), the oldest open clusters may be   
viewed from two perspectives with regard the formation of the Galaxy: 
a halo collapse or a continuous accretion and infall of material from the halo on to the 
Galactic disc. 
Either perspective is possible. The first should justify which were the 
original star formation regions that were the origin of the present old 
clusters in the outer disc and how they travelled there from their place
of  origin.  
The second perspective needs  to test the infall of matter from the halo  
as well as the existence of star systems in the halo. 
Further improvements on these cluster searches and better numbers will give 
us a hint concerning these questions on the origin of the Galaxy. 
A better determination of $C_2$ in the bulge region will tell us  
about the age of bulge clusters if these exist. In this article we have 
observed a relative absence of correlation in the bulge that is somewhat less than the
prediction of our simple model, but  at best the prediction could say, 
as in the case of the anticentre, whether the correlation is greater or 
less than the improved model and enable us to reach further conclusions. 
\subsection*{Acknowledgments}
We thank the anonymous referees for helpful comments that have substantially
improved the content and presentation of this paper.

\begin{appendix} 
 
\section{ Calculations for two intersecting spheres} 
 
The centres of two spheres are separated by a distance $x$, and 
$S_{\rm ss}(x;y,R_{\rm cl})$ is the area of the shell of the first 
sphere (the one 
with radius $y$) contained in the second sphere (Betancort-Rijo 1995),
with radius $R_{\rm cl}$ (see Fig. \ref{Fig:twocircles}).  
There are various cases: 
 
\begin{enumerate} 
 
\item  $R_{\rm cl}>y$: 
 
\begin{enumerate} 
 
\item $x<(R_{\rm cl}-y)$:  the first 
sphere is contained wholly within the second sphere, so that the area of the 
shell that is inside 
the second sphere is the area of the whole shell, and 
 
\begin{equation} 
S_{\rm ss}(x;y,R_{\rm cl})=4\pi y^2 
.\label{Sss1}\end{equation} 
 
\item $(R_{\rm cl}-y)<x<(R_{\rm cl}+y)$: this is the area of the shell for 
$\theta $ up to $\theta _0$, according to Fig. \ref{Fig:twocircles};  
simple trigonometry gives the value of $\theta _0$ (as 
a function of $x$, $y$ and $R_{\rm cl}$) as 
 
\begin{equation} 
R_{\rm cl}^2=x^2+y^2-2xy\cos \theta_0 
\end{equation} 
\begin{equation} 
\cos \theta _0=\frac{x^2+y^2-R_{\rm cl}^2}{-2xy}, 
\label{costheta0}\end{equation} 
\noindent  
and the area of the shell up to $\theta _0$ is 
 
\[ 
S_{\rm ss}(x;y,R_{\rm cl})=2\pi y^2\int _0^{\theta _0}d\theta \ \sin \theta
\]\begin{equation}= 
2\pi y^2(1-\cos \theta _0)
,\end{equation} 
so that, changing $\theta _0$ by the value given in (\ref{costheta0}), we get: 
 
\begin{equation} 
S_{\rm ss}(x;y,R_{\rm cl})=\frac{\pi y}{x}(R_{\rm cl}^2-(x-y)^2) 
.\label{Sss2}\end{equation} 
 
\end{enumerate} 
 
\item  $R_{\rm cl}<y$: 
 
\begin{enumerate} 
 
\item $x<(y-R_{\rm cl})$: here there is no contact between the shell and the 
second 
sphere, so 
 
\begin{equation} 
S_{\rm ss}(x;y,R_{\rm cl})=0 
.\label{Sss3}\end{equation} 
 
\item $(y-R_{\rm cl})<x<(y+R_{\rm cl})$: again, the calculation proceeds as for (\ref{Sss2}): 
 
\begin{equation} 
S_{\rm ss}(x;y,R_{\rm cl})=\frac{\pi y}{x}(R_{\rm cl}^2-(x-y)^2) 
.\label{Sss4}\end{equation} 
 
\end{enumerate} 
\end{enumerate} 

Quantities that we are interested for calculating are
$\int_0 ^{R_{\rm cl}}dx\ x^2 S_{\rm ss}(x;y,R_{\rm cl})$ and 
$\int_{R_{\rm cl}}^{R_{\rm cl}+y}dx$ $\times x^2 S_{\rm ss}(x;y,R_{\rm cl})$. Again, we 
distinguish 
several cases: 
 
\begin{enumerate} 
 
\item $y<R_{\rm cl}$: following (\ref{Sss1}) and (\ref{Sss2}), 
 
\[ 
\int _0^{R_{\rm cl}}dx\ x^2 S_{\rm ss}(x;y,R_{\rm cl})= 
\int _0^{R_{\rm cl}-y}dx\ x^2 4\pi y^2 
\]\[ + \int _{R_{\rm cl}-y}^{R_{\rm cl}} 
dx\ x^2 \frac{\pi y}{x}(R_{\rm cl}^2-(x-y)^2)
\]\begin{equation}=
\frac{4}{3}\pi R_{\rm cl}^3y^2\left (1-\frac{3}{4}\frac{y}{R_{\rm cl}}+ 
\frac{1}{16}\left (\frac{y}{R_{\rm cl}}\right )^3 \right ) 
\end{equation} 
and
 \[ 
\int_{R_{\rm cl}}^{R_{\rm cl}+y}dx\ x^2 S_{\rm ss}(x;y,R_{\rm cl})
\]\[ = 
\int _{R_{\rm cl}}^{R_{\rm cl}+y} 
dx\ x^2 \frac{\pi y}{x}(R_{\rm cl}^2-(x-y)^2)
\]\begin{equation}=
\frac{4}{3}\pi R_{\rm cl}^3y^2\left (\frac{3}{4}\frac{y}{R_{\rm cl}}- 
\frac{1}{16}\left (\frac{y}{R_{\rm cl}}\right )^3 \right ) 
.\end{equation} 
 
\item $R_{\rm cl}<y<2R_{\rm cl}$: following (\ref{Sss3}) and (\ref{Sss4}), 
 
\[ 
\int _0^{R_{\rm cl}}dx\ x^2 S_{\rm ss}(x;y,R_{\rm cl})
\]\[ = 
\int _{y-R_{\rm cl}}^{R_{\rm cl}} 
dx\ x^2 \frac{\pi y}{x}(R_{\rm cl}^2-(x-y)^2)
\]\begin{equation}= 
\frac{4}{3}\pi R_{\rm cl}^3y^2\left (1-\frac{3}{4}\frac{y}{R_{\rm cl}}+ 
\frac{1}{16}\left (\frac{y}{R_{\rm cl}}\right )^3 \right ) 
\end{equation}
and 
 \[ 
\int_{R_{\rm cl}}^{R_{\rm cl}+y}dx\ x^2 S_{\rm ss}(x;y,R_{\rm cl})
\]\[ = 
\int _{R_{\rm cl}}^{R_{\rm cl}+y} 
dx\ x^2 \frac{\pi y}{x}(R_{\rm cl}^2-(x-y)^2)
\]\begin{equation}=
\frac{4}{3}\pi R_{\rm cl}^3y^2\left (\frac{3}{4}\frac{y}{R_{\rm cl}}- 
\frac{1}{16}\left (\frac{y}{R_{\rm cl}}\right )^3 \right ) 
.\end{equation} 
 
\item $y>2R_{\rm cl}$: again, with (\ref{Sss3}) and (\ref{Sss4}), 
 
\begin{equation} 
\int _0^{R_{\rm cl}}dx\ x^2 S_{ss}(x;y,R_{\rm cl})=0 
\end{equation} 
and 
\[
\int_{R_{\rm cl}}^{R_{\rm cl}+y}dx\ x^2 S_{\rm ss}(x;y,R_{\rm cl})
\]\begin{equation}= 
\int _{y-R_{\rm cl}}^{R_{\rm cl}+y} 
dx\ x^2 \frac{\pi y}{x}(R_{\rm cl}^2-(x-y)^2)= 
\frac{4}{3}\pi R_{\rm cl}^3y^2 
.\end{equation} 
 
\end{enumerate}

\section{Edge effects in the measurement of the TPACF 
in rectangular fields} 
 
We have a rectangular surface with size $a\times b$ ($x$ from $0$ to $a$ and 
$y$ from $0$ to $b$). A ring of negligible thickness and of radius 
$\theta $ whose centre is  
located at $(x,y)$ contains  part of the surface inside the rectangle, 
$f(x,y)$, and the rest of it is outside the rectangle. Due to the  
loss of a part of the ring surface outside the rectangle we  measure only a 
fraction $F_{\rm BE}$ 
of the star counts  separated by an angular distance $\theta $. Assuming 
that the distribution of the stars in the rectangle is homogeneous, we have 
 
\begin{equation} 
F_{\rm BE}=\frac{1}{a\ b}\int_0^adx\int_0^bdy f(x,y) 
.\end{equation} 
 
The value of $f(x,y)$ depends on the case. With the condition  
$2\theta <a$, $2\theta <b$ (which is to be satisfied in the case 
considered here), 
the only posibilities are: 
 
\begin{enumerate} 
 
\item when $\theta <x<(a-\theta)$ and $\theta <y<(b-\theta)$: the whole 
ring is inside the rectangle, so 
 
\begin{equation} 
f(x,y)=1 
.\end{equation} 
 
\item when $\theta <x<(a-\theta)$ and $y<\theta$: the portion of  
ring contained inside the rectangle is from the angle $-\sin^{-1}(y/\theta)$ 
to $\pi+\sin ^{-1}(y/\theta)$ (the value of $\sin ^{-1}(\cdots)$ between  
0 and $\pi /2$). Then, 
 
\begin{equation} 
f(x,y)=\frac{\pi +2\sin^{-1}(y/\theta)}{2\pi}  
.\end{equation} 
 
\item when $x<\theta$ and $\theta <y<(b-\theta)$: similarly to the previous 
case: 
 
\begin{equation} 
f(x,y)=\frac{\pi +2\sin^{-1}(x/\theta)}{2\pi}  
.\end{equation} 
 
\item when $x<\theta$, $y<\theta $ and $(x^2+y^2)<\theta ^2$: the 
portion of ring inside the rectangle is from the angle 
$-\sin^{-1}(y/\theta)$ to $\pi/2+\sin^{-1}(x/\theta)$, so 
 
\begin{equation} 
f(x,y)=\frac{\pi/2+\sin^{-1}(x/\theta)+\sin^{-1}(y/\theta)}{2\pi}  
.\end{equation} 
 
\item when $x<\theta$, $y<\theta $ and $(x^2+y^2)>\theta ^2$: this case   
is similar to the previous one, but we must add another portion of ring 
which is between $-(\pi /2+\sin^{-1}(x/\theta))$ and  
$\pi +\sin^{-1}(y/\theta)$, so 
 
\begin{equation} 
f(x,y)=\frac{2\sin^{-1}(x/\theta)+2\sin^{-1}(y/\theta)}{2\pi}  
.\end{equation} 
 
\end{enumerate} 
 
Other possible situations are avoided by the symmetry properties of the  
integral 
used in the evaluation of $F_{\rm BE}$, which is reduced to the following 
calculation: 
 
\[ 
F_{\rm BE}(\theta )=\frac{1}{a\ b}[(a-2\theta)(b-2\theta)]
\]\[ + 
\frac{2}{a\ b}\left[\int_\theta ^{a-\theta}dx\int_0^\theta dy 
\frac{\pi+2\sin^{-1}(y/r)}{2\pi }\right]
\]\[ +\frac{2}{a\ b} 
\left[\int_\theta ^{b-\theta}dy\int_0^\theta dx 
\frac{\pi+2\sin^{-1}(x/r)}{2\pi }\right]
\]\[+
\frac{4}{a\ b}\left[\int_0 ^{\theta }dx\int_0^\theta dy 
\frac{\pi /2+\sin^{-1}(x/r)+\sin^{-1}(y/r)}{2\pi }\right]
\]\[+
\frac{4}{a\ b}\left[\int_0 ^{\theta }dx\int_{\sqrt{\theta ^2-x^2}}^\theta dy 
\frac{-\pi /2+\sin^{-1}(x/r)+\sin^{-1}(y/r)}{2\pi }\right] 
\]\begin{equation} 
=1-\frac{2}{\pi }\left(\frac{1}{a}+\frac{1}{b}\right)\theta+ 
\left(\frac{9}{\pi}+\frac{\pi }{2}-4\right) \frac{1}{a\ b} \theta ^2 
.\end{equation} 
 
\end{appendix}

\end{document}